\documentclass[10pt,journal,compsoc]{IEEEtran}
\usepackage[pdftex]{graphicx}
\usepackage{epstopdf}
\usepackage{times}
\usepackage{epsfig}
\usepackage{graphicx}
\usepackage{amsmath}
\usepackage{amssymb}
\usepackage{url}
\usepackage{array}
\usepackage{booktabs}
\usepackage{multirow}
\usepackage{subfigure}
\usepackage{array}
\usepackage{xcolor}
\usepackage{float}
\usepackage{mathrsfs,enumitem}
\usepackage{hyperref}
\usepackage{ulem}
\usepackage{cite}
\usepackage{mathtools}
\usepackage{ulem} 
\usepackage{hyperref}
\hypersetup{pdfborder={0 0 0}, colorlinks=false,linkcolor=blue}

\newcommand{\mX}{\ensuremath{\mathbf{X}}}

\newif\ifannotated

\annotatedtrue

\ifannotated

\newcommand{\delete}[1]{{\color{red}{\sout{#1}}}}

\newcommand{\margincomment}[1]{\marginpar{$\Rightarrow$\color{red}\fbox{\parbox{\linewidth}{\color{black}\scriptsize#1}}}}
\else

\newcommand{\delete}[1]{{\ignorespaces}}

\newcommand{\margincomment}[1]{}
\fi

\begin{document}
%
\title{A Dual Camera System for High Spatiotemporal Resolution Video Acquisition}


\author{Ming Cheng, Zhan Ma, M. Salman Asif, Yiling Xu, Haojie Liu, Wenbo Bao, and Jun Sun
\thanks{M. Cheng, Z. Ma and H. Liu are with Nanjing University, Nanjing, Jiangsu, China. M. S. Asif is with the University of California at Riverside. M. Cheng is also with Shanghai JiaoTong University, Shanghai, China. Y. Xu, W. Bao and J. Sun are with Shanghai JiaoTong University, Shanghai, China. {\it Z. Ma is the corresponding author of this paper. M. S. Asif and Y. Xu are co-corresponding authors of this paper.}}
\thanks{This paper is supported in part by National Natural Science Foundation of China (61971282), National Key Research and Development Project of China Science and Technology Exchange Center (2018YFE0206700) and Scientific Research Plan of the Science and Technology Commission of Shanghai Municipality (18511105402).}
}

%




\IEEEtitleabstractindextext{%
\begin{abstract}
This paper presents a dual camera system for high spatiotemporal resolution (HSTR) video acquisition, where one camera shoots a video with high spatial resolution and low frame rate (HSR-LFR) and another one captures a low spatial resolution and high frame rate (LSR-HFR) video. 
Our main goal is to combine videos from LSR-HFR and HSR-LFR cameras to create an HSTR video. We propose an end-to-end learning framework, {AWnet}, mainly consisting of a FlowNet and a FusionNet that learn an adaptive weighting function in pixel domain to combine inputs in a frame recurrent fashion. To improve the reconstruction quality for cameras used in reality, we also introduce noise regularization under the same framework.
 Our method has demonstrated noticeable performance gains in terms of both objective PSNR measurement in simulation with different publicly available video and light-field datasets and subjective evaluation with real data captured by dual iPhone 7 and Grasshopper3 cameras.  Ablation studies are further conducted to investigate and explore various aspects (such as reference structure, camera parallax, exposure time, etc) of our system to fully understand its capability for potential applications.
\end{abstract}

}

\maketitle

\IEEEdisplaynontitleabstractindextext

%
\IEEEpeerreviewmaketitle

\section{Introduction}


%
%
%
%


High-speed  cameras play an important role in various modern imaging and photography tasks including sports photography, film special effects, scientific research, and industrial monitoring. They allow us to see very fast phenomena that are easily overlooked and can not be captured at ordinary speed, such as a droplet, full-speed fan rotation, or even a gun fire. These cameras can capture videos at high frame-rates that range from several hundred to  several thousand frames per second (FPS), while an ordinary camera operates at 30 to 60 FPS.
The high frame-rates often come at the expense of spatial resolution; especially, consumer-level cameras that sacrifice the spatial resolution to maintain the high frame rate acquisition. For example, popular iPhone 7 can capture 4K videos at 30 FPS, but can only offer 720p resolution at 240 FPS because of the limitation of the data I/O throughput.
Some special-purpose and professional high-speed cameras can capture high spatiotemporal resolution (HSTR) videos, but they are typically very expensive  (e.g., Phantom Flex4K\footnote{\url{https://www.phantomhighspeed.com/products/cameras/4kmedia/flex4k}} with price starting at \$110K) and beyond the budget of a majority of consumers.

\begin{figure}[htbp]
    \centering
    \includegraphics[width=0.36\textwidth]{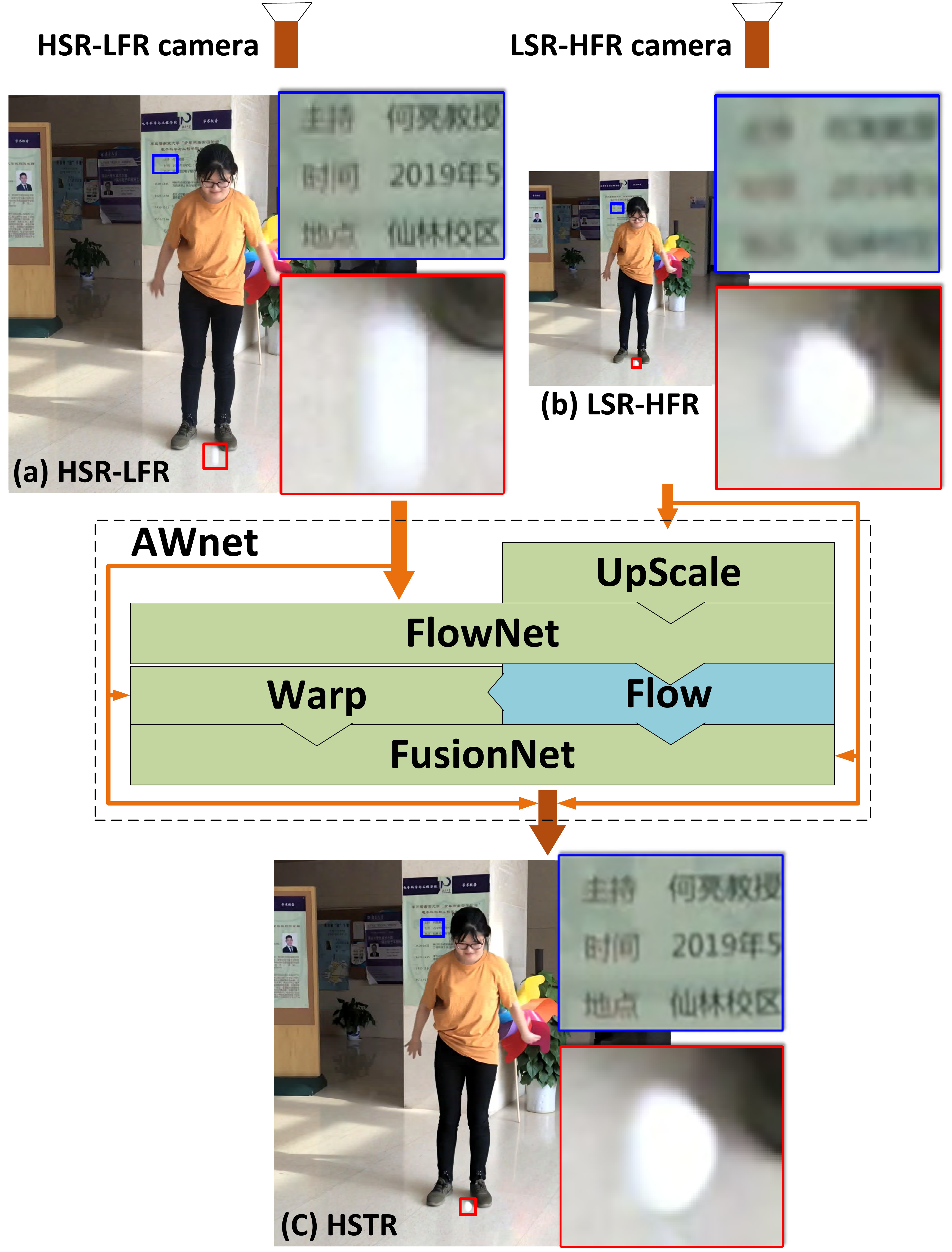}
\caption{Snapshots of high spatial resolution-low frame rate (HSR-LFR) and low spatial resolution-high frame rate (LSR-HFR) videos and synthesized high spatiotemporal resolution (HSTR) video. A woman is throwing a ping-pong ball in indoor space. (a) HSR-LFR video 4K@30FPS frame with zoomed-in region showing motion blur; (b) LSR-HFR video 720p@240FPS frame with zoomed-region showing spatial blur; (c) HSTR video 4K@240FPS frame.}
\label{fig:HSR_vs_LSR_capture}
\end{figure}


A na\"ive solution to obtain a HSTR video from a video with high spatial resolution and low frame rate (HSR-LFR) or low spatial resolution and high frame rate (LSR-HFR) is to upsample along temporal or spatial direction, respectively. Upsampling in temporal resolution or frame rate upconversion of an HSR-LFR video (e.g., 4K at 30FPS) involves imputing missing frames by interpolating motion between the observed frames, which is challenging because of the motion blur introduced by long exposure and inaccuracies in motion representation under the commonly-used uniform translational motion assumption~\cite{xue2019video}.
On the other hand, upsampling spatial resolution of a LSR-HFR video can be performed using a variety of existing super-resolution methodologies~\cite{lim2017enhanced,zheng2018crossnet}, but they often provide smoothed images in which high frequency spatial details of the captured scene are missing.
Figure~\ref{fig:HSR_vs_LSR_capture} highlights these effects, where HSR-LFR (4K at 30FPS) video contains motion blur in the regions of fast motion and LSR-HFR (720p at 240FPS) has a uniform spatial blur because of limited spatial resolution.

\begin{figure*}[htbp]
\centering
\subfigure[]{\includegraphics[scale=0.72]{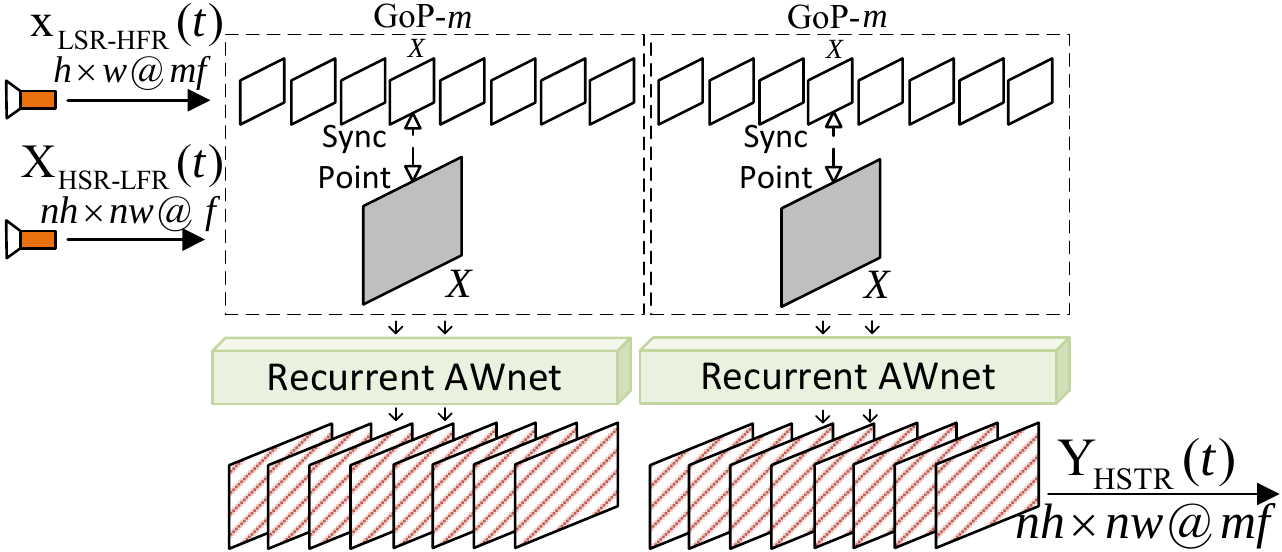}
\label{sfig:RefSR_GoP_Proc}}
\subfigure[]{\includegraphics[scale=0.75]{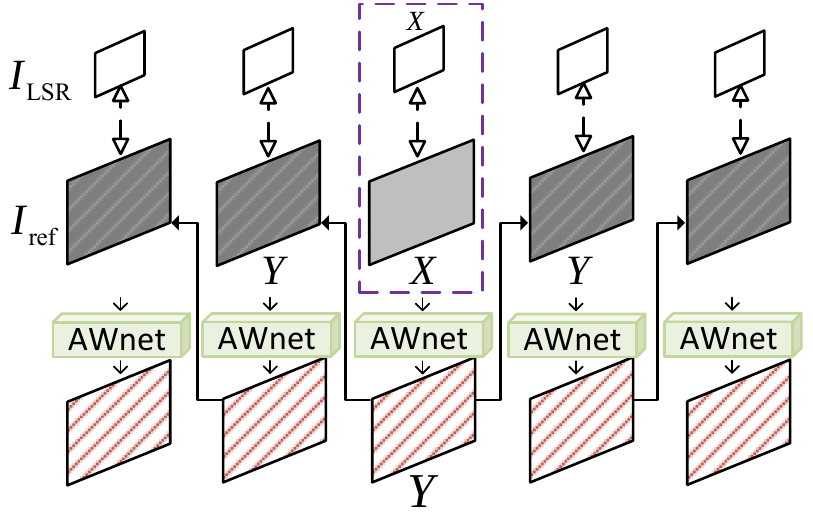}
\label{sfig:RefSR_recurrent}}
\subfigure[]{\includegraphics[scale=0.68]{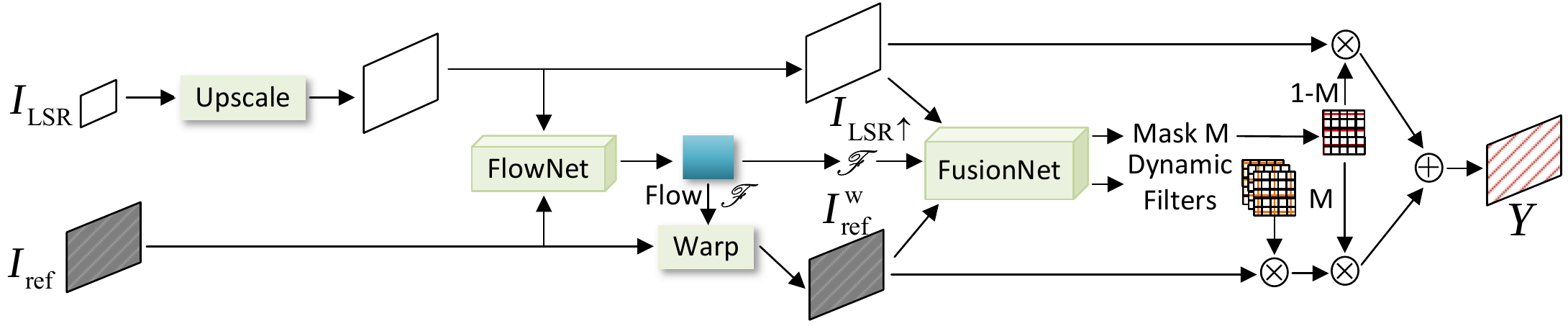} \label{sfig:refSR}}
\subfigure[]{\includegraphics[scale=0.63]{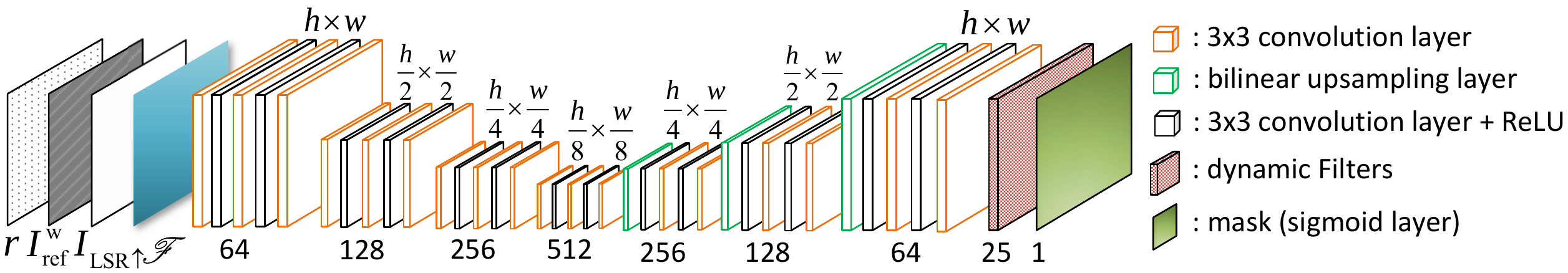}\label{sfig:FusionNet}}
\caption{\textbf{A Dual Camera System for High Spatiotemporal Acquisition:} (a) dual camera setup with one LSR-HFR video capture (e.g., ${\bf x}_{\tt LSR-HFR}(t)$ with $h\times w$ at $mf$ FPS), the other HSR-LFR video shooting (e.g., ${\bf X}_{\tt HSR-LFR}(t)$ with $nh\times nw$ at $f$ FPS) and synthesized HSTR video (e.g., ${\bf Y}_{\tt HSTR}(t)$ with $nh\times nw$ at $mf$ FPS); (b) Recurrent RefSR structure for  $Y$ synthesis using $I_{\tt LSR}$ and $I_{\tt ref}$ at each time instant; (c) Proposed AWnet for dual camera input with cascaded FlowNet and FusionNet to learn adaptive weights for final synthesis; (d) An U-net style~\cite{ronneberger2015u} FusionNet structure for dynamic filter and mask generation. $\bigoplus$ and $\bigotimes$ are element-wise addition and multiplication. 
}
\label{fig:RefSR_framework}
\end{figure*}

In this paper, we propose a dual camera system for HSTR video ${\bf Y}_{\tt HSTR}(t)$ acquisition, as shown in Fig.~\ref{fig:HSR_vs_LSR_capture}, where one camera captures a HSR-LFR video ${\bf X}_{\tt HSR-LFR}(t)$ with rich spatial information (i.e., sharp spatial details for textures and edges), and the other one records a HFR-LSR video ${\bf x}_{\tt LSR-HFR}(t)$ with fine-grain temporal information (i.e., intricate motion flows). We then fuse these two videos via a learning-based approach to produce a final HSTR video with both appealing spatial details and accurate motion. In another words, we aim to transfer the rich spatial details from HSR-LFR frame to the associated LSR-HFR frames while retaining accurate motion in the entire sequence.

Our method performs spatiotemporal super-resolution in a frame recurrent manner within a synchronized GoP (group of pictures) to exploit the spatial-temporal priors in Fig.~\ref{sfig:RefSR_GoP_Proc}. A detailed block diagram of our proposed method is shown in Fig.~\ref{sfig:RefSR_recurrent}. The synthesis process has two main parts: FlowNet and FusionNet, which are placed consecutively in Fig.~\ref{sfig:refSR}.
$I_{\tt LSR}$ denotes a frame captured with LSR-HFR camera. The FlowNet accepts an upsampled LSR-HFR image  ($I_{\tt LSR\uparrow}$) and a reference image ($I_{\tt ref}$) to provide the optical flow (denoted as $\mathscr{F}$) and a warped reference image ($I_{\tt ref}^w$).
The reference image can either be a frame from the HSR-LFR camera at a synchronized time instant or a synthesized frame $Y$. %
Our dual camera based system can be viewed as a method in the class of "super-resolution with reference" (RefSR) methods~\cite{zheng2018crossnet,boominathan2014improving}.
The FusionNet accepts the optical flow, upsampled LSR-HFR image, and warped reference image, and learns dynamic filters and masking pattern that are used to adaptively weigh the contribution of $I_{\tt LSR}$  and $I_{\tt ref}$ for a high-quality reconstruction $Y$. We use PWC-Net~\cite{sun2018pwc} as our FlowNet and a U-net as our FusionNet~\cite{ronneberger2015u}. More details about network architecture are provided in Section~\ref{sec:dual_camera_system} and in Fig.~\ref{sfig:FusionNet}. Our method learns adaptive weights to combine the hybrid inputs, therefore, we refer to it as an adaptive weighting network (AWnet).

Our proposed AWnet is trained using Vimeo90K dataset. We first evaluate the performance of our method using simulations on publicly available datasets. Then we measure the performance of our method on videos captured with our custom dual-camera prototype. In our experimental evaluations, we observe that the quality of HSTR video degrades when we directly use models trained with Vimeo90K training images.
One reason for the performance degradation is the presence of large sensor noise in  $I_{\tt LSR}$ when LSR-HFR video is captured with short exposure time (especially under low light conditions). Vimeo90K training data is virtually free of noise and other nonidealities that a real data capture encounters. To make our system robust to noise, we introduce noise at various levels in original Vimeo90K training data when performing the end-to-end learning. Such noise regularization can intelligently shift weights  between $I_{\tt LSR}$ and $I_{\tt ref}$, offering much better reconstruction quality, under the same framework.

{{Extensive simulations are conducted using both simulation data from publicly accessible videos, dual-camera captures, and light field datasets (such as Vimeo90K~\cite{xue2019video}, KITTI~\cite{menze2015object}, Flower~\cite{srinivasan2017learning}, LFVideo~\cite{wang2017light} and Stanford Light Field~\cite{StanfordLF} datasets\footnote{Note that these datasets are widely used in literature for performance benchmark~\cite{zheng2018crossnet}.})}}, and real data (captured with custom-built dual iPhone 7 or Grasshopper3 cameras). Our proposed AWnet demonstrates noticeable performance gains over the existing super-resolution and frame interpolation methods, in objective and subjective measures. In our tests with simulations using Vimeo90K testing samples, our proposed model offers $\sim0.7$ dB PSNR gain compared to the state-of-the-art CrossNet~\cite{zheng2018crossnet} and $\sim 6$ dB PSNR compared to the popular single-image super-resolution (SISR) method EDSR~\cite{lim2017enhanced}. Our proposed AWnet provides the best performance on other video and lightfield datasets.
In our tests with real data, we observe perceptual enhancements for various scenarios with indoor and outdoor activities under different lighting conditions.

We also offer ablation studies to fully understand the capability of our dual camera AWnet system, by analyzing various aspects in practice, such as the impacts of upscaling filters, reference structure, camera parallaxes, exposure time, etc. All these tests demonstrate the efficiency of our dual camera system for super-resolution and frame interpolation, to maintain sharp spatial details and accurate temporal motions jointly, leading to the state-of-the-art performance.

Main contributions of this work are highlighted below.

\begin{itemize}
    \item A practical system for high spatiotemporal video acquisition uses a dual off-the-shelf camera setup. Videos from two cameras, operating at different spatial and temporal resolution, are combined using an end-to-end learning-based adaptive weighting to preserve spatial and temporal information in both inputs for a high-quality reconstruction.
    \item Cascaded FlowNet and FusionNet are applied to learn embedded spatial and temporal features for adaptive weights derivation in a frame recurrent way. These weights can be regularized using added noise to efficiently handle noise and other nonidealities in real data captured with consumer cameras.
    \item Our dual camera AWnet system demonstrates the state-of-the-art performance for super-resolution and frame interpolation, using both simulation data from public and real data captured by cameras.

    \item We analyze the robustness and efficiency of our system through a series of ablation studies to explore the impacts of upscaling filters, reference structure, camera parallaxes, exposure time, etc, which promises generalization in a variety of practical scenarios.
\end{itemize}


The remainder of this paper is structured as follows. Section~\ref{sec:related_work} provides a brief overview of related work in literature, including system prototypes and applications. Section~\ref{sec:dual_camera_system} details our proposed system and associated learning algorithms, followed by training processing in Section~\ref{sec:train}. The experimental results on simulation data and real data captured by cameras are shown in Section~\ref{sec:exp}. We further break down our system to analyze and study its various aspects, such as the camera parallax, scaling filters, etc, in Section~\ref{sec:ablation_studies}. Finally, conclusion is drawn in Section~\ref{sec:conclusion}. Table~\ref{tab:notation} contains a list of all the notions and acronyms used throughout this paper.

\begin{table}[htbp]
\centering
\caption{Notations and Abbreviations}
\label{tab:notation}
\renewcommand
\arraystretch{1.05}
\begin{tabular}{c|c}
\hline
    Abbr. & Description  \\
\hline
    HSTR & High Spatiotemporal Resolution\\
    HFR & High Frame Rate (or Temporal Resolution)\\
    LFR & Low Frame Rate (or Temporal Resolution)\\
    HSR & High Spatial Resolution (or Frame Size)\\
    LSR & Low  Spatial Resolution (or Frame Size)\\
    \hline

    ${\bf Y}_{\tt HSTR}(t)$ & Output HSTR Video\\
    $Y_{t_i} = {\bf Y}_{\tt HSTR}(t_i)$ & A Frame of HSTR Video at time $t_i$\\
    ${\bf x}_{\tt LSR-HFR}(t)$ & Input LSR-HFR Video\\
    $x_{t_i} = {\bf x}_{\tt LSR-HFR}(t_i)$ & A Frame of LSR-HFR Video at time $t_i$\\
    ${\bf X}_{\tt HSR-LFR}(t)$ & Input HSR-LFR Video\\
    $X_{t_i} = {\bf X}_{\tt HSR-LFR}(t_i)$ &  A Frame of HSR-LFR Video at time  $t_i$\\
    ${\bf\bar{X}}_{\tt LSR-HFR}(t)$ & Upscaled ${\bf x}_{\tt LSR-HFR}(t)$\\
    $I_{\tt ref}$ & A Frame from either ${\bf X}_{\tt HSR-LFR}(t)$ or  ${\bf Y}_{\tt HSTR}(t)$ \\
    $I_{\tt LSR}$ & A Frame from  ${\bf x}_{\tt LSR-HFR}(t)$\\
    $I_{\tt LSR\uparrow}$ & A Frame from ${\bf\bar{X}}_{\tt LSR-HFR}(t)$\\
    $I_{\tt ref}^{w}$ & warped  $I_{\tt ref}$\\
    \hline
    $h$, $w$ & Height \& width of $x_{t_i}$\\
    $f$ & frame rate of ${\bf X}_{\tt HSR-LFR}(t)$\\
    $n$, $m$ & scaling factor of respective SR \& FR\\
    \hline
    SISR & Single-Image Super Resolution\\
    RefSR & Super Resolution with Reference\\
    SNR & Signal-to-Noise Ratio\\
    PSNR & Peak Signal-to-Noise Ratio\\
    SSIM & Structural Similarity\\
\hline
\end{tabular}
\end{table}

\section{Related Work} \label{sec:related_work}
This work is closely related to the high-speed video acquisition, multi-camera (or array) system, super-resolution, and frame interpolation. we will briefly review these topics below.

\textbf{High-speed Cameras.} High-speed (or slow-motion) video capturing has been widely used these days. For example, iPhone Xs Max can capture a slow-motion video with 1080p at 240FPS, and Samsung Galaxy S10 offers 720p at 960FPS for a short time (0.2s). Consumer mobile devices often sacrifice spatial resolution and other image quality-related factors to ensure the high frame rate throughput. Professional high-speed cameras can support both high spatial and temporal resolution, typically coming along with a bulky body and an inconvenient price. Professional i-SPEED 726 can shoot 2K at 8512FPS and 1080p at 12742FPS, but its price range is above USD100,000. In recent years, we have noticed the increasing adoptions of multi-camera design from professional device to mobiles, such as dual camera iPhone or four camera Huawei P30. Thus, we have new opportunities to combine inputs from hybrid cameras operating at different speeds and synthesize high spatiotemporal resolution video.

\textbf{Multi-Camera System.} In pursuit of ultra-high spatial resolution, such as gigapixel, multi-camera systems and arrays have been developed~\cite{brady2012multiscale,parallel_camera,lu2019high}. In particular, cameras with different properties have been combined in a hybrid setup to sample and synthesize images from a variety of light components, such as hyperspectral imaging~\cite{cao2016computational}, low light imaging~\cite{lowlight_imaging}, and light fields~\cite{wu2017light,zheng2018crossnet}.
Pelican imaging and Light are two recent companies that launched products with multiple cameras on board capturing a diverse set of images and synthesizing a desired image with high dynamic range, long-range depth, or  light field \cite{wu2017light} .
The hybrid camera or multi-camera systems have also been used to combine different imaging sources to perform super-resolution and frame interpolation (for frame rate up-conversion)~\cite{zheng2018crossnet}.
Our proposed work also belongs to the hybrid camera setup that captures two video streams, one at HSR-LFR and the other one at LSR-HFR, and combine them to synthesize a final HSTR video.

\textbf{Super-Resolution.} Single image super resolution (SISR) methods upscale individual images, which include traditional methods based on bilinear and bicubic filters and recently-introduced learning-based techniques~\cite{ledig2017photo,lim2017enhanced}. SISR methods can be easily extended to support video or multi-frame super-resolution~\cite{caballero2017real,sajjadi2018frame,jo2018deep}. In the case of multi-camera setup, super-resolution can be performed using some source as a reference \cite{freeman2002example,boominathan2014improving}. Low resolution images/videos can be upscaled with references (e.g., RefSR) from other viewpoints, leading to significant quality improvement~\cite{boominathan2014improving,zheng2017combining,zheng2017learning,zheng2018crossnet}.
Such RefSR approaches have also been widely used in lightfield imaging~\cite{boominathan2014improving,zheng2018crossnet}.

\textbf{Frame Interpolation.} Linear translation motion is a conventional assumption that has been extensively used in different interpolation-based methods to impute missing frames for frame rate up-conversion. Motion estimation can be performed using classical block-based or dense optical flow-based methods~\cite{baker2011database,niklaus2018context,xue2019video,jiang2018super,bao2018high,bao2018memc,DAIN}. Classical optimization-based and modern learning-based methods mainly try to retain smooth motion along the temporal trajectory while resolving occlusion-induced artifacts. Accurate motion flow estimation remains a challenging task because of the inconsistent object movements and motion-induced occlusion. As we discussed below, this issue can be significantly alleviated with the help of a high frame-rate video as a reference.



\section{Dual Camera System for High Spatiotemporal Resolution Video} \label{sec:dual_camera_system}

We propose a dual camera system for HSTR video acquisition, as illustrated in Fig.~\ref{fig:RefSR_framework}. One camera records a HSR-LFR video ${\bf X}_{\tt HSR-LFR}(t)$ with $nh\times{nw}$ at $f$ FPS, while the other one captures a LSR-HFR video ${\bf x}_{\tt LSR-HFR}(t)$ with $h\times{w}$ at $mf$ FPS. We learn an adaptive model to weigh contributions from the two input videos and synthesize a final  HSTR  output video ${\bf Y}_{\tt HSTR}(t)$ with $nh\times{nw}$ at $mf$ FPS. We use integer multipliers, $m$ and $n$, for simplicity in this work, but different multipliers can be easily used. In subsequent sections, we first offer experimental observations that a single camera setup could not provide high-quality reconstruction of HSTR video via na\"ive  spatial super-revolution or temporal frame interpolation. Then we discuss our dual camera system and algorithm development. Note that even though we particularly emphasize current work in a dual camera setup, this work can be generalized to other multi-camera configurations since the RefSR structure can be flexibly extended.

\subsection{Single Camera System}
Let us consider the following model for an image frame captured at time instance $t$ of a camera as
\begin{equation}
\label{equ:real_camera}
I(t) = \int^{t}_{t-T}S(\tau)d\tau+n(t),
\end{equation}
where $S(\tau)$ is the instantaneous photon density reflected from the physical scene, $T$ denotes the exposure time, and $n(t)$ is the noise accumulated in the camera during a single exposure and the subsequent readout process. In other words, image is represented as the accumulated photons during the exposure time (according to the shutter speed). {A typical consumer camera used in mobile devices\footnote{We use mobile phone camera as an example for its massive market adoption.} usually automatically adjusts the aperture size, ISO settings, and shutter speed according to a specific ``shooting mode''. Thus, exposure time duration $T$ is uncertain and changes according to the scene content.}

 For example, normal video capture in iPhone 7 offers HSR (e.g., 4K) at $f=$ 30FPS (i.e., HSR-LFR video), leading to rich spatial details but blurred motion. On the other hand, the slow-motion mode provides LSR (e.g., $<$ 720p) at $mf$ = 240FPS (LSR-HFR video), resulting in accurate motion acquisition at the expense of spatial information, dynamic range, and SNR. Figure~\ref{fig:HSR_vs_LSR_capture} shows two snapshots for respective HSR-LFR and LSR-HFR videos. As we can see, the spatial quality of  the LSR-HFR frame is poor because some spatial information is missing in the slow-motion mode. Similarly, motion blur is clearly observed for the HSR-LFR frame because of insufficient temporal sampling.

Our experimental results in Tables~\ref{table:objective_perf_sim}, \ref{table:FI} and Figs.~\ref{fig:indoor_m_light},~\ref{fig:outdoor_low_light},~\ref{fig:frame_interp_pingpong} show that frame-by-frame super-resolution or  temporal interpolation of frames from a single camera could not provide the same quality as the dual camera setup, both objectively and subjectively.

These observations suggest that we are not capable of reconstructing high-quality HSTR video by applying spatial interpolation on the LSR-HFR video or temporal interpolation on HSR-LFR video. Therefore, we propose to use a dual camera system in which the main problem is to synthesize two input videos while preserving the sharp spatial details from the HSR-LFR video and real motions from the LSR-HFR video.

\subsection{Dual Camera System} \label{ssec:dual_camera}

For a dual camera system setup, given a LSR-HFR video ${\bf x}_{\tt LSR-HFR}(t)$ recorded at $mf$ FPS and an additional HSR-LFR video ${\bf X}_{\tt HSR-LFR}(t)$ recorded at $f$ FPS from another view, we wish to generate a final HSTR video ${\bf Y}_{\tt HSTR}(t)$ at $mf$ FPS.
To recover high-quality HSTR video, we seek to preserve the real motion field from the LSR-HFR camera input and the detailed spatial information from the HSR-LSR input. To do that, we need to design an effective mechanism to extract, transfer, and fuse appropriate information or features intelligently from the two inputs.

{Dual cameras are configured and synchronized with a certain baseline distance to capture the instantaneous frames of the same scene. Because of the different frame rates for the respective HSR-LFR and LSR-HFR cameras, the exposure time of the HSR-LFR frame are often much longer than the LSR-HFR frames, especially in the low-light settings. Based on the imaging model in \eqref{equ:real_camera}, we can also consider the HSR-LFR frame as the high-spatial-resolution version of the summation of associated LSR-HFR frames, which will model the motion blur.}


We divide the entire processing into two major steps: (1) optical flow estimation to compensate for motion and parallaxes to facilitate image fusion; (2) fusion processing to compute appropriate weighting functions (e.g., dynamic filter and mask in our work) through extensive feature learning.
We propose to apply learned networks to perform aforementioned temporal and spatial feature extraction, transfer, and fusion. For convenience, we note them as ``FlowNet'' and ``FusionNet'' respectively, shown in Fig.~\ref{sfig:refSR}.

We will synthesize the HSTR video in batches of group-of-pictures (GoP), as shown in Fig.~\ref{sfig:RefSR_GoP_Proc}. Because the same processing pattern is applied in each GoP, we will explain the specific steps for a single GoP. A GoP is a set of frames from the LSR-HFR video that are aligned to a specific time stamp of the HSR-LFR video. Since we assume that the LSR-HFR video is recorded at  $mf$ FPS and the HSR-LFR at $f$ FPS, we find it convenient to use a GoP with $m$ frames that we call GoP-$m$ in the remainder of this paper. A GoP-$m$ consists of $m$ LSR-HFR frames that are aligned with one HSR-LFR frame. In particular, we assume that the HSR-LFR frame timestamp is in the middle of the timestamps for all the LSR-HFR frames in a GoP-$m$. For instance, a GoP-$m$ that contains LSR-HFR frames at times $[t - k' \Delta t, \ldots, t + k\Delta t]$ are synchronized with HSR-LFR frame at $t$, where $k' = \lceil \frac{m}{2} \rceil-1$, $ k = \lfloor \frac{m}{2}\rfloor$, and $\Delta t$ is the time interval between two adjacent HSR-LFR frames. To maximize the use of spatial information from synchronized HSR-LFR frames, we use the synchronized HSR-LFR frame at $t$ to first super-resolve its synchronized LSR-HFR frame at time $t$ and then super-resolve its adjacent $m-1$ frames in a frame-recurrent manner.

\subsubsection{Updating Synchronization Frame}
We start with the synchronization frame for every HSR-LFR video frame. Let us denote an HSR-LFR frame at time $t_i$ as $X_{t_i} = \mX_{\tt HSR-LFR}(t_i)$.
We upscale the LSR-HFR video to the same spatial size as HSR-LFR, i.e.,
\begin{align}
    {\bf\bar{X}}(t) = \mathcal{U}({\bf x}_{\tt LSR-HFR}(t)),
\end{align}
where $\mathcal{U}(\cdot)$ denotes a spatial upscaling operator that either performs bilinear/bicubic interpolation or some other type of SISR methods (e.g., EDSR~\cite{lim2017enhanced}).
%
%
Let us denote the upscaled LSR-HFR frames as  $\bar{X}_{t_i} = \bar \mX(t_i)$.
At synchronization time $t_i$, we use the ${\bar{X}}_{t_i}$ and ${X}_{t_i}$ to produce an HSTR frame $Y_{t_i} = {\bf Y}_{\tt HSTR}(t_i)$, as shown in Fig.~\ref{sfig:RefSR_recurrent}.

The remaining HSTR frames in the GoP-$m$, are then generated in a frame-recurrent manner to fully exploit and leverage temporal priors of reconstructions. In other words, we first create super-resolved version of the synchronization frame and then reconstruct one HSTR frame at a time using its immediate super-resolved neighbor.
The GoP-$m$ centered at timestamp $t_i$ can be written as $[\bar{X}_{t_i - k' \Delta t}, \ldots, \bar{X}_{t_i + k\Delta t}]$, where $k' = \lceil \frac{m}{2} \rceil-1$, $ k = \lfloor \frac{m}{2}\rfloor$, and $\Delta t = (t_{i+1} - t_{i}) / m$. Thus, we start at the center and estimate $Y_{t_i+ \Delta t}$ and $Y_{t_i-\Delta t}$ using $Y_{t_i}$; then we estimate $Y_{t_i+k \Delta t}$ using $Y_{t_i + (k-1)\Delta t}$ and $Y_{t_i-k' \Delta t}$ using $Y_{t_i - (k'-1)\Delta t}$ for all the frames in the GoP-$m$.

The entire process described above is analogous to a RefSR method, where the reference is a high spatial resolution frame either from a snapshot captured by a HSR-LFR camera  or from a synthesized HSTR frame.
We use a frame-recurrent method for super-resolution because the motion between two adjacent frames in LSR-HFR video is often small and the estimates are reliable, which provides a robust recovery. In comparison, motion estimates between the synchronization frame and all the other frames in a GoP-$m$ can be large and unreliable, which seriously affects the performance of RefSR~\cite{zheng2018crossnet}.

\subsubsection{FlowNet}
A popular approach to obtain the temporal motion fields or features is by using the {\it optical flow} \cite{burton1978thinking}. Let us assume that we can compute optical flow between two frames $I_{\tt ref}, I_{\tt LSR\uparrow}$ as
\begin{align}
    \mathscr{F} = {\sf FlowNet}\left(I_{\tt ref}, I_{\tt LSR\uparrow}\right),
\end{align}
where $I_{\tt ref}$ refers to a reference frame and $I_{\tt LSR\uparrow}$
denote the upscaled LSR-HFR frame at any specific time stamp.
For the synchronization timestamp $t_i$, $I_{\tt LSR\uparrow} = {\bar{X}}_{t_i}$ and $I_{\tt ref} = X_{t_i}$. For the frames at timestamp $t_i+ k\Delta t$, $I_{\tt LSR\uparrow}= {\bar{X}}_{t= t_i + k\Delta t}$ and $I_{\tt ref} ={Y}_{t_i + (k-1)\Delta t}$ with $ 1\leq k\leq\lfloor m/2\rfloor $. Similarly for the frames at timestamps $ t_i - k'\Delta t$, $I_{\tt LSR\uparrow} = {\bar{\bf X}}_{\tt LSR-HFR}(t_i -k'\Delta t)$ and $I_{\tt ref} = {Y}_{t_i - (k'-1)\Delta t }$ with $1\leq k'\leq\lceil m/2\rceil-1$. 

A number of deep neural networks have been proposed to deal with the optical flow~\cite{niklaus2018context,FlowNet2,sun2018pwc}. We adopt a pretrained optical flow model PWC-Net~\cite{sun2018pwc} as the FlowNet in our frame recurrent AWnet, and then use our data to fine-tune the model through retraining. Pretraining the optical flow network with mass labeled data greatly improves the convergence speed and accuracy of the flow calculation in our work.
The size of the estimated optical flow by PWC-Net is $\frac{1}{4}$th of the input image along both spatial dimensions. We use a simple bilinear upsampling method to upscale the low-resolution optical flow field to the same size of the input images.


Spatial details of the reference frame can be transferred using extracted optical flow through the warping operation. Let us denote the warped reference image as $I_{\tt ref}^{w}$. Since the optical flow is upsampled using a bilinear filter, it often leads to an over-smoothed output.

\subsubsection{FusionNet}
To preserve the fine motion details, we borrow the idea of dynamic filtering to refine our flow.
{{Dynamic filtering for motion estimation and motion compensation has been recently used in \cite{jia2016dynamic}.
{{It estimates an independent convolution kernel for each pixel, which can correctly describe the motion behaviors of each pixel individually. It especially shows accurate estimation and compensation of small motions, but it is not as effective as global optical flow  for estimating large motion because of the limited size of convolutional filters. Thus we use dynamic filters to complement the optical flow devised in FlowNet for better performance. As far as we know, we are the first to combine flow network for flow estimation and dynamic filter network for motion refinement.}}}}

We observe in our experiments that even with the refinement of the optical flow using dynamic filters, the warped  $I_{\tt ref}^w$ fails in region with occlusions and suffers from motion and warping artifacts. In such regions, we need the information from $I_{\tt LSR\uparrow}$. Therefore, we learn a mask to create a weighted combination of the warped reference image and $I_{\tt LSR\uparrow}$ for every pixel. Our FusionNet is designed to perform motion refinement using dynamic filters and provide an adaptive weighting mask as an output. The structure of FusionNet is illustrated in Fig.~\ref{sfig:FusionNet}.

To utilize all the information in the available frames, we explicitly feed warped reference frame $I_{\tt ref}^{w} = {\sf Warp}(I_{\tt ref}, \mathscr{F})$, upscaled LSR-HFR frame $I_{\tt LSR\uparrow}$, optical flow $\mathscr{F}$, residual between warped reference and upscaled LSR-HFR frame $r=I_{\tt ref}^{w} - I_{\tt LSR\uparrow}$, to the FusionNet as inputs. We can describe the FusionNet as the following function:
\begin{align}
{\bf Fm} = {\sf FusionNet}(I_{\tt ref}^{w}, I_{\tt LSR\uparrow}, \mathscr{F} , r), \label{eq:fusionnet}
\end{align}
whose output has 26 channels that we use to calculate the dynamic filter and adaptive weighting mask.

We use the popular U-net architecture~\cite{ronneberger2015u} for our FusionNet, as shown in Fig.~\ref{sfig:FusionNet}. We downscale the  feature maps by a factor of two in each of the three downscaling layers and then upscale the features using bilinear interpolation for computational efficiency. In contrast, existing approaches use transposed convolution layers for upscaling, which is computationally expensive and also causes some checkerboard artifacts. We do not use any skip connection in conventional U-net, which greatly reduces the memory requirements of our network. The output of FusionNet is a three-dimensional tensor of feature maps that has 26-channels and same spatial size as the input image frame. We use the first 25 channels to produce $5\times5$ dynamic filters (one filter per pixel), and the last one to produce the weighted mask for every pixel. Let us denote the dynamic filter for pixel $(x,y)$ as a $5\times 5$ $K_{x,y}$ matrix that can be written as
\begin{equation}
    K_{x,y}(i,j) = {\bf Fm}(x,y,5(i-1)+j-1), \quad \text{for } i,j = 1,\dots,5.
\end{equation}
Let us denote the weighted mask for the entire image as a matrix $M$ with same size as input image whose value at pixel $(x,y)$ can be written as
\begin{equation}
{ M}(x,y) = {\sf sigmoid}[{\bf Fm}(x,y,25)]. \label{eq:adapt_weighting_mask}
\end{equation}

To summarize, an output frame $Y$ of the reconstructed HSTR  video can be synthesized for pixel $(x,y)$ as
\begin{align}
Y(x,y) = {M}(x,y)&I_{\tt ref}^{wk}(x,y) \nonumber\\&+ {(1- M(x,y))}I_{\tt LSR\uparrow}(x,y),  \label{eq:output_Y}
\end{align}
where\begin{align}
I_{\tt ref}^{wk}(x,y)=\sum_{i,j=1}^5{K}_{x,y}(i,j)I_{\tt ref}^{w}(x-3+i,y-3+j).\label{equ:dynamic_filter}
\end{align}
The reconstructed $Y$ in \eqref{eq:output_Y} (together with $I_{\tt LSR\uparrow}$ from next time instant) will be fed into our AWnet module (as a typical RefSR) in a recurrent manner to recover other HSTR frames reconstruction as exemplified in Fig.~\ref{sfig:RefSR_recurrent}.

\subsubsection{Reference Structure}
The synthesis model discussed in the previous sections assumes an  ultra-low latency application scenario. Thus, we use a single reference frame in AWnet and denote it as a {\it Single-Reference} AWnet. In practice, we can easily extend this method to include multiple references. One obvious example is to use two references such that one reference frame precedes current LSR-HFR frame, and the other one succeeds it (e.g., two consecutive HSR-LFR frames in Fig.~\ref{sfig:RefSR_GoP_Proc} used to super-resolve $I_{\tt LSR}$s in between ). We refer to it as the {\it Multi-Reference} AWnet.

To enable {\it Multi-Reference} AWnet with two references, we use two FlowNets to estimate the flows between two pairs: ($I_{\tt ref0}$, $I_{\tt LSR\uparrow}$) and ($I_{\tt ref1}$, $I_{\tt LSR\uparrow}$). Then we warp $I_{\tt ref0}$ and $I_{\tt ref1}$ to compute  $I_{\tt ref0}^w$ and $I_{\tt ref1}^w$ accordingly. Similar to the {\it Single-Reference} AWnet, we input warped reference frames, upscaled LSR-HFR frame, optical flows, and residual frames between warped reference and upscaled LSR-HFR frame to the same FusionNet for dynamic filters and masks generation. In the case of Two-Reference AWnet, we increase the number of output channels in the FusionNet from 26 to 53, where the first 50 produce two sets of $5\times 5$ dynamic filters $K^{\tt ref0}$ and $K^{\tt ref1}$, and the remaining 3 provide adaptive weighting masks $M_{\tt ref0}$, $M_{\tt ref1}$ and $M_{\tt LSR\uparrow}$. We replace the original {\sf sigmoid} function in  \eqref{eq:adapt_weighting_mask} with a {\sf softmax} function to enable the multi-reference weighting, where the sum of these three masks at each pixel position is 1. Finally, reconstructed $Y(x,y)$ with two reference frames can be expressed as
\begin{align}
Y(x,y) = &{M_{\tt ref0}}(x,y)I_{\tt ref0}^{wk}(x,y)+{M_{\tt ref1}}(x,y)I_{\tt ref1}^{wk}(x,y) \nonumber\\
&\mbox{~~~~~~~~~~~}+ M_{\tt LSR\uparrow}(x,y)I_{\tt LSR\uparrow}(x,y).
\end{align}
In this example, we can set original $I_{\tt ref}$ in Fig.~\ref{sfig:refSR} as $I_{\tt ref0}$, and duplicate a reference branch for incoming $I_{\tt ref1}$.

We can further extend two-reference AWnet to support more reference frames by duplicating reference branches in Fig.~\ref{sfig:refSR} and modifying the last layer of the FusionNet in Fig.~\ref{sfig:FusionNet} for dynamic filters and mask generation appropriately. Note that {\sf softmax} activation can be utilized to support multiple reference weighting.

We will show in the ablation studies below that multiple-reference AWnet provides noticeable improvement over single-reference AWNet in the synthesized reconstruction with better spatial texture details and temporal continuity.
Objective improvement measured by averaged PSNR over thousands of videos in Vimeo90K test dataset is given in Table~\ref{table:reference-x}, and subjective comparisons are supplemented with the demonstration videos at our website\footnote{\url{http://yun.nju.edu.cn/d/def5ea7074/?p=/MultiReferenceAWnet&mode=list}}.  Applying either {\it Single-Reference} or {\it Multi-Reference} AWnet is dependent on the underlying application, where {\it Single-Reference} is preferred if ultra-low latency condition is demanded, otherwise, {\it Multi-Reference} AWnet is favored.

\section{Training}\label{sec:train}

\subsection{Training Dataset}\label{sec:dataset}
We use the Vimeo90K dataset~\cite{xue2019video} to train our model. The Vimeo90K dataset has 64,612 septuplets for training, where each septuplet contains 7 consecutive video frames at a size of $256\times{448}$ pixels. For each septuplet, we randomly select two consecutive frames as a pair for training. Specifically, one frame is used as the {\it reference} $I_{\tt ref}$ to mimic the input image from a HSR-LFR camera, and a downscaled version of the next frame is used as the {\it target frame}. We apply a native bicubic downsampling filter offered by the open source FFmpeg\footnote{\url{www.ffmpeg.org}} to mimic the input image $I_{\tt LSR}$ from a LSR-HFR camera. And, we randomly crop each frame from its original resolution to a size of $256\times{384}$ on-the-fly for training data augmentation.

\subsection{Training Strategy and Loss Function}\label{Sec:training}
Training process for our network has four main steps. We use the Adam~\cite{kingma2014adam} optimizer by setting its parameters $\beta_{1}$ and $\beta_{2}$ to 0.9 and 0.999, respectively. We use a batch size of 4. Details of every training step are as follows.
\begin{itemize}
    \item \textbf{Step 0: FlowNet Initialization.} We use the pretrained PWC-Net~\cite{sun2018pwc} to initialize our FlowNet, which is trained with a large set of data with ground truth optical flow. The inputs of PWC-Net are two consecutive frames at the same resolution.
    \item \textbf{Step 1: FlowNet Fine-tuning.} The reference frame $I_{\tt ref}$ and  low-resolution target frame $I_{\tt LSR}$ have a large gap in their sizes. Thus, we implement a fine-tuning step to improve the FlowNet. First, we upscale $I_{\tt LSR}$ to $I_{\tt LSR\uparrow}$ with the same size as $I_{\tt ref}$. Then we compute optical flow between $I_{\tt LSR\uparrow}$ and $I_{\tt ref}$ using PWC-Net. The computed optical flow $\mathscr{F}$ is then used to warp $I_{\tt ref}$ and produce $I_{\tt ref}^w$. Then we apply an $\ell_1$ norm-based warping loss to fine-tune the FlowNet, which is shown below:
    \begin{equation}
        \mathcal{L}_{warp} = ||I_{\tt gt}-I_{\tt ref}^w||_{1},
    \end{equation} where $I_{\tt gt}$ is the high-resolution ground truth of $I_{\tt LSR}$. A small learning rate of $1e-6$ is used to fine-tune the FlowNet with 40k iterations. A similar loss function has been used in \cite{jaderberg2015spatial}.

    \item \textbf{Step 2: FusionNet Pretraining.} A pretraining step is also used for FusionNet. To train the FusionNet, we fix the FlowNet and let the network select appropriate parameters for FusionNet during training. We use an $\ell_1$ loss between the output $Y$ and the ground truth $I_{\tt gt}$, given as
    \begin{equation}\label{equ:l1loss}
        \mathcal{L}_{rec} = ||I_{\tt gt}- Y||_{1}.
    \end{equation}
    We set the learning rate to $1e-4$ and train the network for 100k iterations, according to our extensive simulation studies.

    \item \textbf{Step 3: End-to-End Joint Training.} Starting with our pretrained models, we jointly train FlowNet and FusionNet by minimizing the same end-to-end $\ell_1$ loss in \eqref{equ:l1loss}. In this step, we set learning rate to $10^{-5}$ for FusionNet and $3\times 10^{-6}$ for FlowNet over 100k iterations. With such pre- and joint-training, network model can converge faster with more robust and reliable behavior.
\end{itemize}
All networks are implemented and verified using PyTorch. In subsequent sections, we describe the experiments we performed to evaluate different aspects of proposed AWnet for our dual camera system.
\begin{table}[t]
\caption{Objective Performance Comparison of Super-Resolution Methods on Vimeo90K Dataset~\cite{xue2019video}.}
\label{table:objective_perf_sim}
\centering
\renewcommand
\arraystretch{1.05}
\begin{tabular}{ c c  c  c  c  c }
\specialrule{0em}{1pt}{1pt}
\hline
       \multicolumn{2}{c}{\multirow{2}{*}{Methods}} & \multicolumn{2}{c}{$4\times$} & \multicolumn{2}{c}{$8\times$}\\
\cline{3-6}
       && PSNR  & SSIM  & PSNR  &  SSIM  \\
\hline
      \multirow{2}{*}{SR} & EDSR~\cite{lim2017enhanced} & 33.11 & 0.9413 & 28.20 & 0.8702 \\
      &ToFlow-SR~\cite{xue2019video} & 33.08 & 0.9417  &-& -\\
\hline
      \multirow{3}{*}{RefSR} & PM~\cite{boominathan2014improving} & 35.06  & 0.9670 & 31.30 & 0.9380 \\
      &CrossNet~\cite{zheng2018crossnet} & 39.17 &0.9852 & 36.15 & 0.9766\\
      &{\bf AWnet} & \textbf{39.88} & \textbf{0.9862} & \textbf{36.63} & \textbf{0.9768}\\
\hline
\end{tabular}
\end{table}


\begin{table}[t]
\caption{Objective Performance Comparison of Frame Interpolation using Vimeo90K~\cite{xue2019video} (downscaled 4th frame used as reference in AWnet).}
\label{table:FI}
\centering
\renewcommand
\arraystretch{1.05}
\begin{tabular}{ c  c  c  }
\specialrule{0em}{1pt}{1pt}
\hline
       & PSNR  & SSIM   \\
\hline

      ToFlow-Intp.~\cite{xue2019video} & 33.46 & 0.9615 \\
      {\bf AWnet} with 1/64 reference & \textbf{36.63} & \textbf{0.9768}\\
      {\bf AWnet} with 1/16 reference & \textbf{39.88} & \textbf{0.9862} \\
\hline
\end{tabular}
\end{table}

\section{Experiments}
\label{sec:exp}
We conduct experiments on two types of videos. One type is the ``simulation data'' that has images/videos from the existing and public accessible datasets (e.g., Vimeo90K, KITTI, Flower, LFVideo and Stanford Light Field datasets); the other type is the ``real data'' captured by real cameras (e.g., iPhone 7 and Grasshopper3 cameras) under different settings.

\subsection{Performance Comparison using Simulation Data} \label{sec:perf_comp_sim_data}
We first compare our method with the state-of-the-art SISR method EDSR~\cite{lim2017enhanced}, task-oriented video super-resolution method ToFlow-SR~\cite{xue2019video}, conventional RefSR patchmatch (PM)~\cite{boominathan2014improving}, and the state-of-the-art learning-based RefSR CrossNet~\cite{zheng2018crossnet}. To be fair, we retrain CrossNet with our dataset following the training strategy suggested in \cite{zheng2018crossnet}. 

{\bf Super-resolution:} We first use the test set with 7,824 septuplets from Vimeo90K~\cite{xue2019video} for performance comparison. We select the fourth image in each septuplet for evaluation following the suggestion in \cite{xue2019video,bao2018memc}. For video super-resolution method, the input is the downscaled septuplet sequence and the target is the super-resolved fourth frame. For a single frame or image RefSR, we downscale the fourth frame and use the fifth frame as the reference frame. The results are presented in Table~\ref{table:objective_perf_sim}.
We use PSNR and Structural Similarity (SSIM)~\cite{wang2004image} as our performance metrics for evaluation.  Results show that our method has superior performance in both PSNR and SSIM for $4\times$ super-resolution along both spatial dimensions. For PSNR, it yields $\approx$ 0.7 dB, 4.8 dB, 6.8 dB, and 6.7 dB gains against CrossNet, PM, ToFlow-SR, and EDSR, respectively. Similar gains are produced for $8\times$ super-resolution factor, demonstrating the generalization of our work to various application scenarios.

{In addition to the experiments using Vimeo90K testing samples, we also tested other datasets such as KITTI, {{Flower, LFVideo}} and Stanford Light Field data to evaluate the performance of our proposed AWnet. We discuss those experiments in Section~\ref{sec:ablation_studies} where we analyze the impact of camera parallax.}


{\bf Frame interpolation:} Our AWnet can also be used to interpolate missing intermediate frames (usually at high spatial resolution) with the help from another LSR-HFR input. Such frame interpolation is also supported by optical flow based methods, such as ToFlow-Intp in~\cite{xue2019video}. We use the third and the fifth frames from the testing septuplets to interpolate missing fourth frame. But for our method, we downscale fourth frame (e.g., $8\times$ resolution downscaling at both spatial dimension) as another input. The results in Table~\ref{table:FI} suggest that even a thumbnail-size image of its original source (e.g., $1/8\times{1/8}$ the size of the original image), can improve the quality of the interpolated intermediate frame significantly. A remarkable 6.4 dB PSNR gain is recorded compared to ToFlow-Intp~\cite{xue2019video} when scaling the fourth image to its $1/4\times{1/4}$ size (i.e., $16\times$ fewer pixels) and 3.2 dB PSNR gains for the case when scaling fourth image to its $1/8\times{1/8}$ size (i.e., $64\times$ fewer pixels).

{\bf Model efficiency:} Our AWnet demands less system resource with less space and time complexity requirements. For example, AWnet model has 109.5 MB parameters, about 25\% reduction when compared with the CrossNet model at a size of 140.8 MB parameters. When upscaling a snapshot at a factor of 8$\times$ spatially to the size of 640$\times$448, AWnet consumes about 0.12 second with 1499 MB running memory (e.g., about 60\% reduction against the running memory consumption of CrossNet), while CrossNet is about 0.18 second with 4511 MB running memory. As an comparative anchor, traditional PM~\cite{boominathan2014improving} uses 55.9 seconds due to iterative patch match.

\subsection{Performance Studies using Real Data}

\begin{figure}[t]
\centering
    \subfigure[]{\includegraphics[scale=0.032]{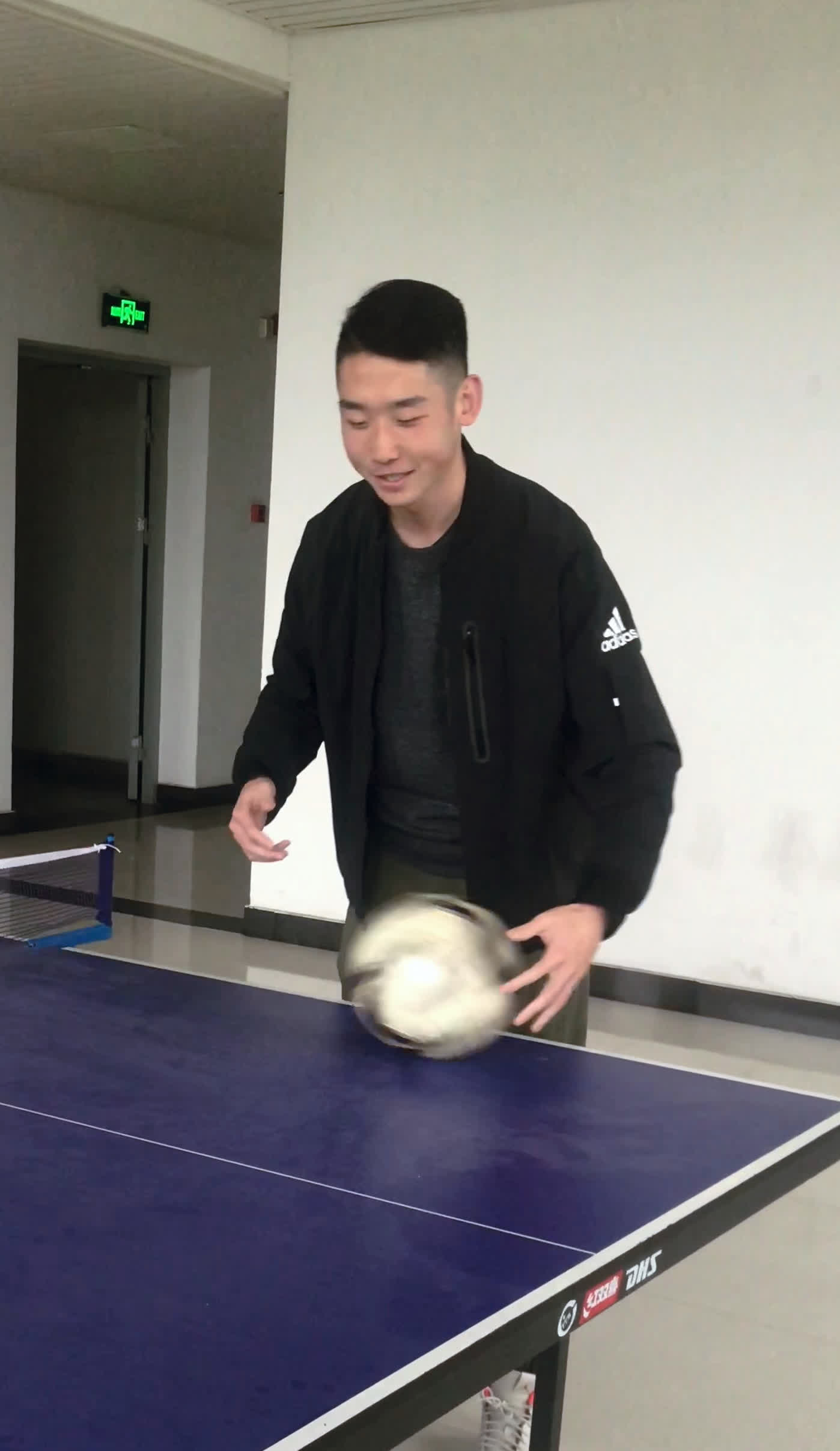}}
    \subfigure[]{\includegraphics[scale=0.032]{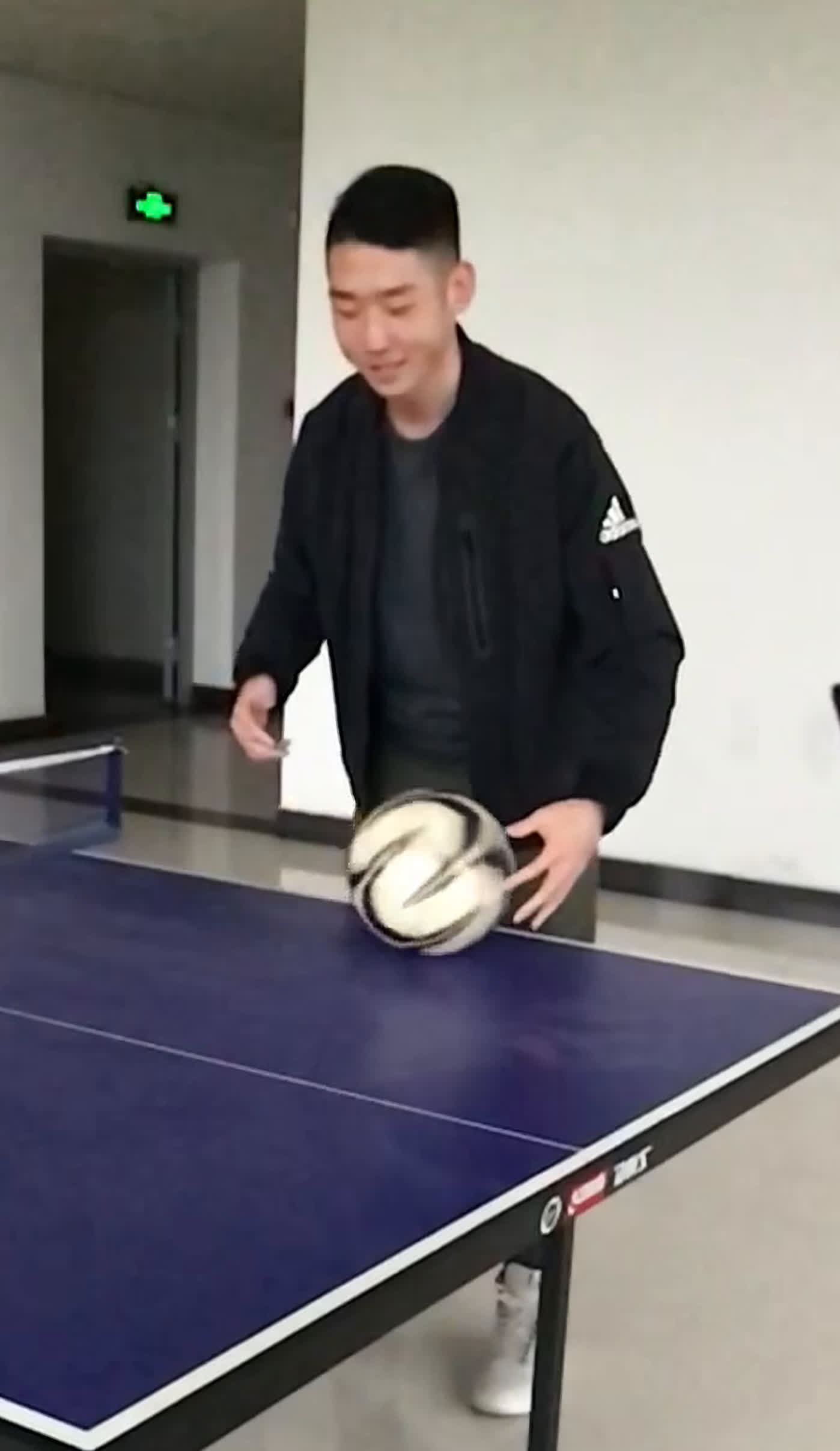}}
    \subfigure[]{\includegraphics[scale=0.032]{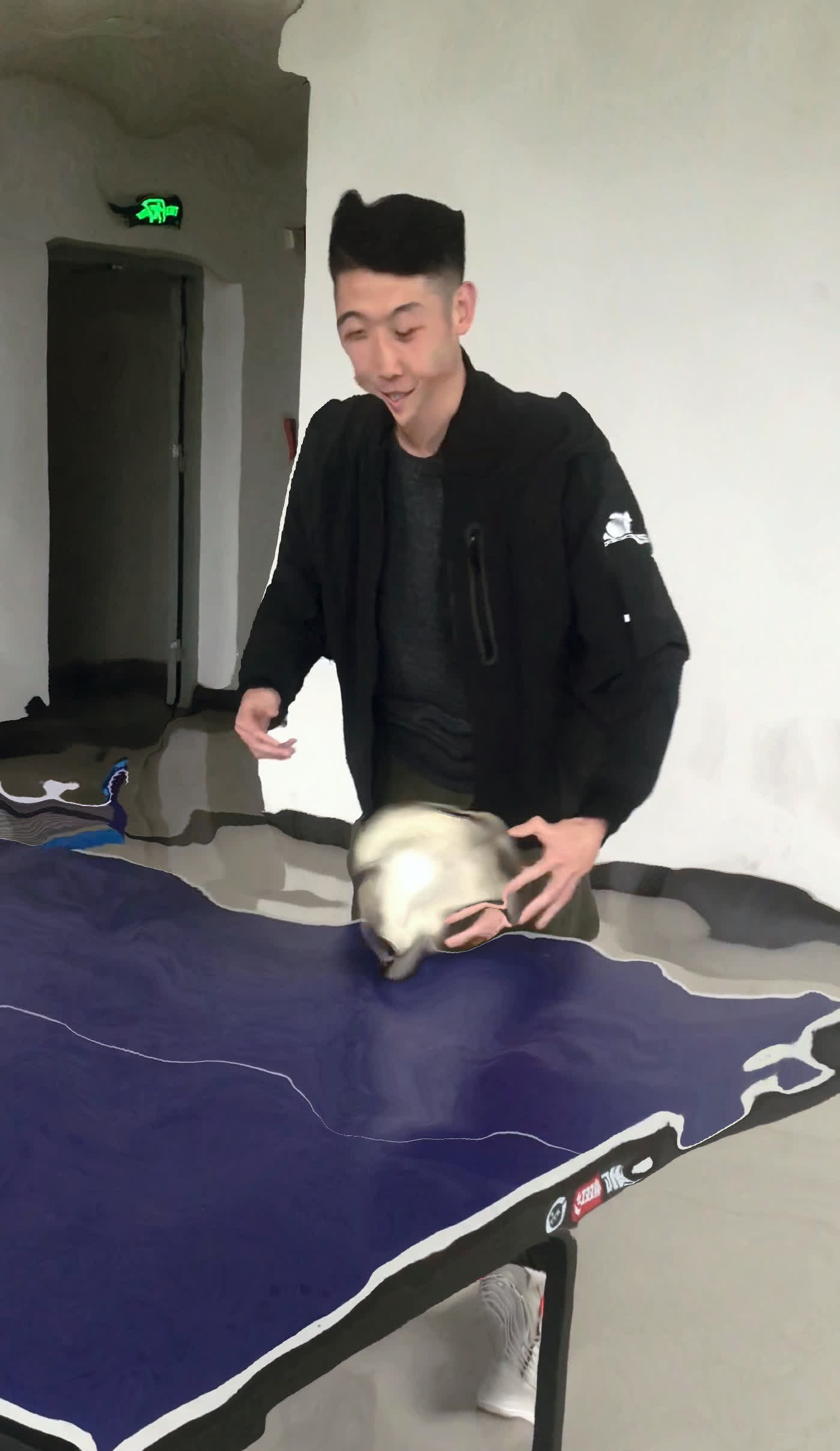}}
    \subfigure[]{\includegraphics[scale=0.032]{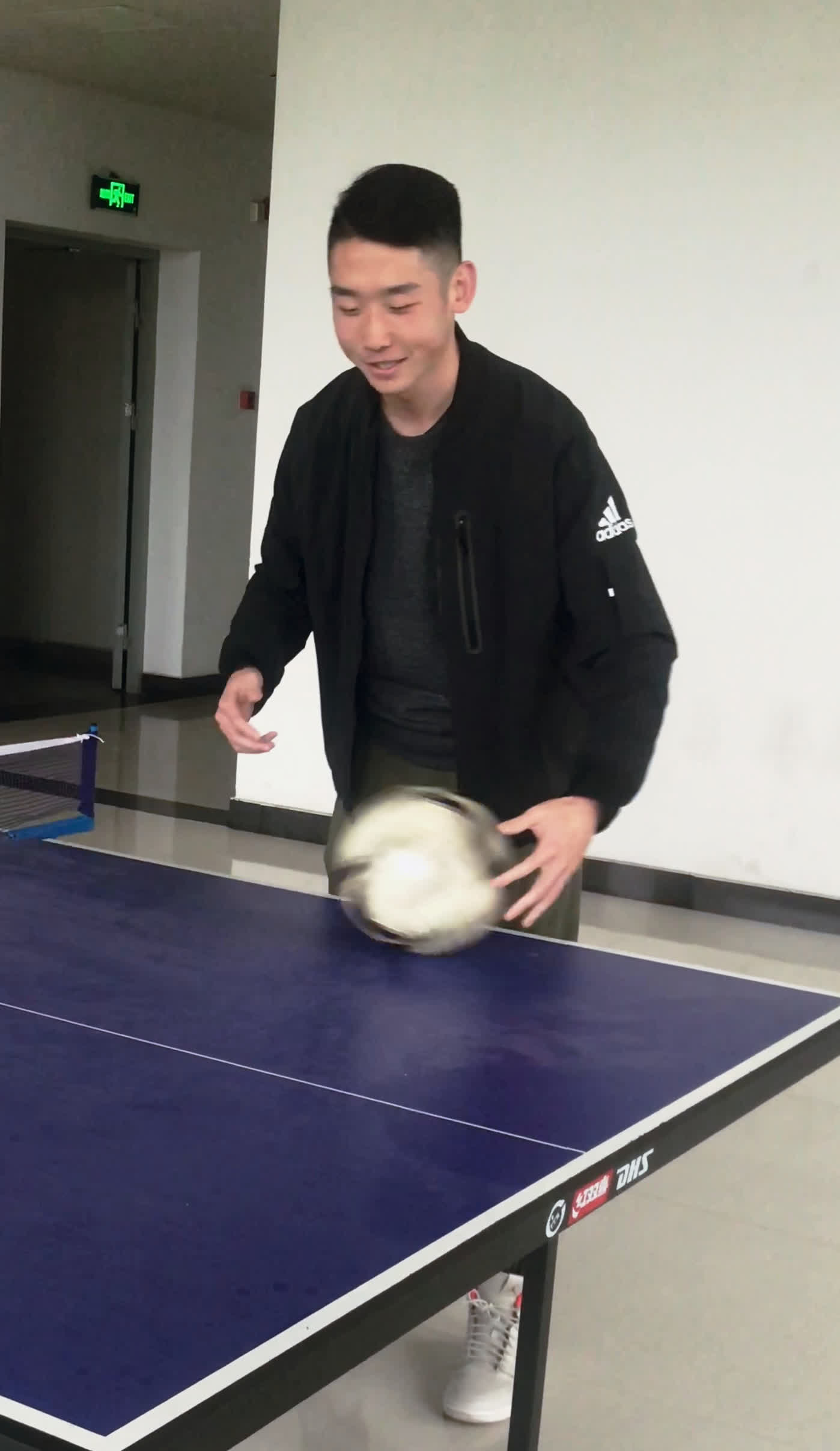}}
    \subfigure[]{\includegraphics[scale=0.032]{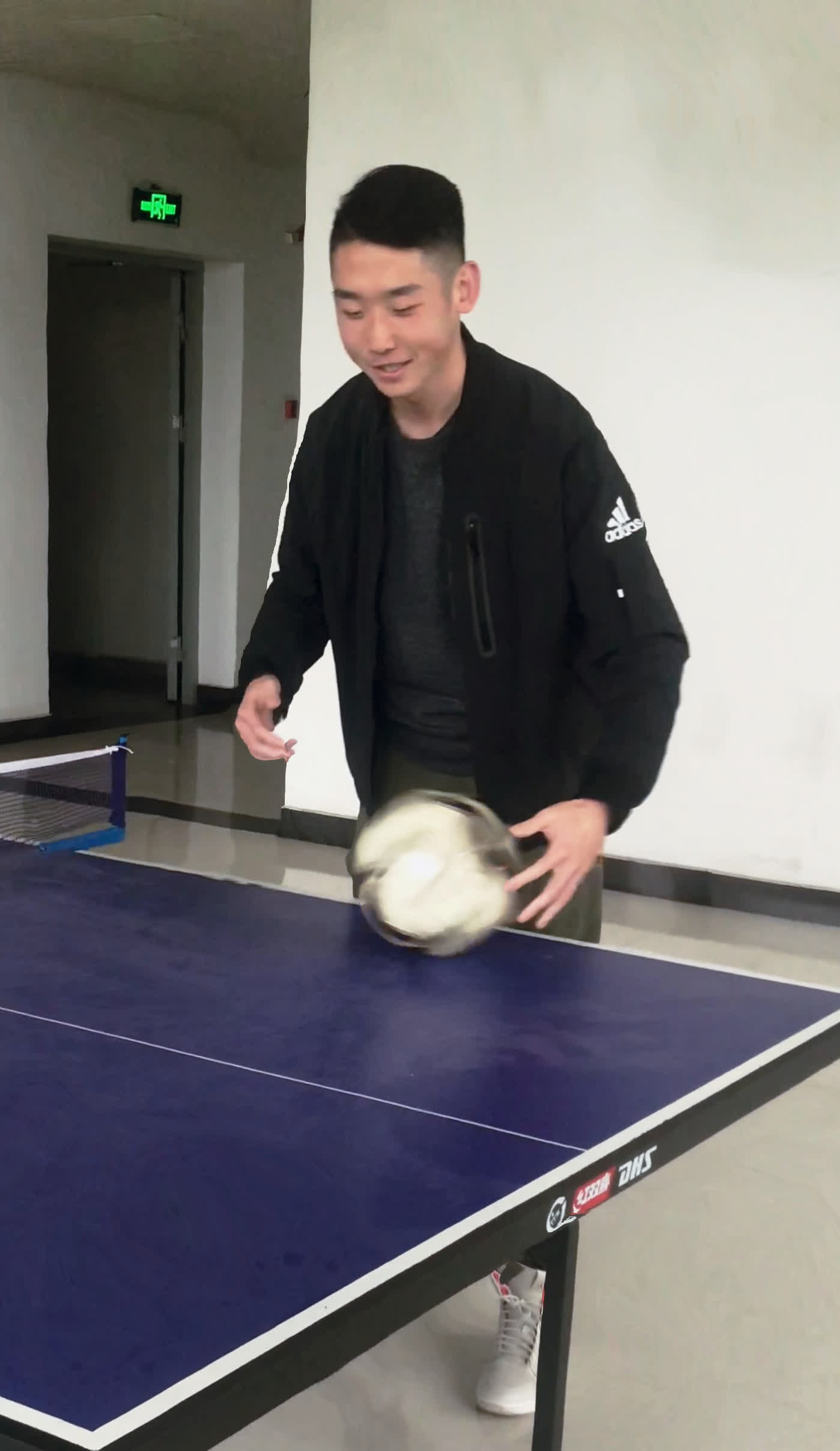}}
\caption{{\bf Dual Camera Alignment.}  (a) HSR-LFR frame $I_{\tt ref}$; (b) LSR-HFR frame $I_{\tt LSR}$; (c) HSR-LFR frame $I_{\tt ref}$ frame warped using optical flow only; (d) HSR-LFR frame $I_{\tt ref}$ warped using mesh-based homography; (e) HSR-LFR frame warped using both mesh-based homography and optical flow. (a) and (b) are captured using dual iPhone 7 with different views.}
\label{fig:mesh}
\end{figure}

\begin{figure}[t]
    \centering
    \includegraphics[width=0.4\textwidth]{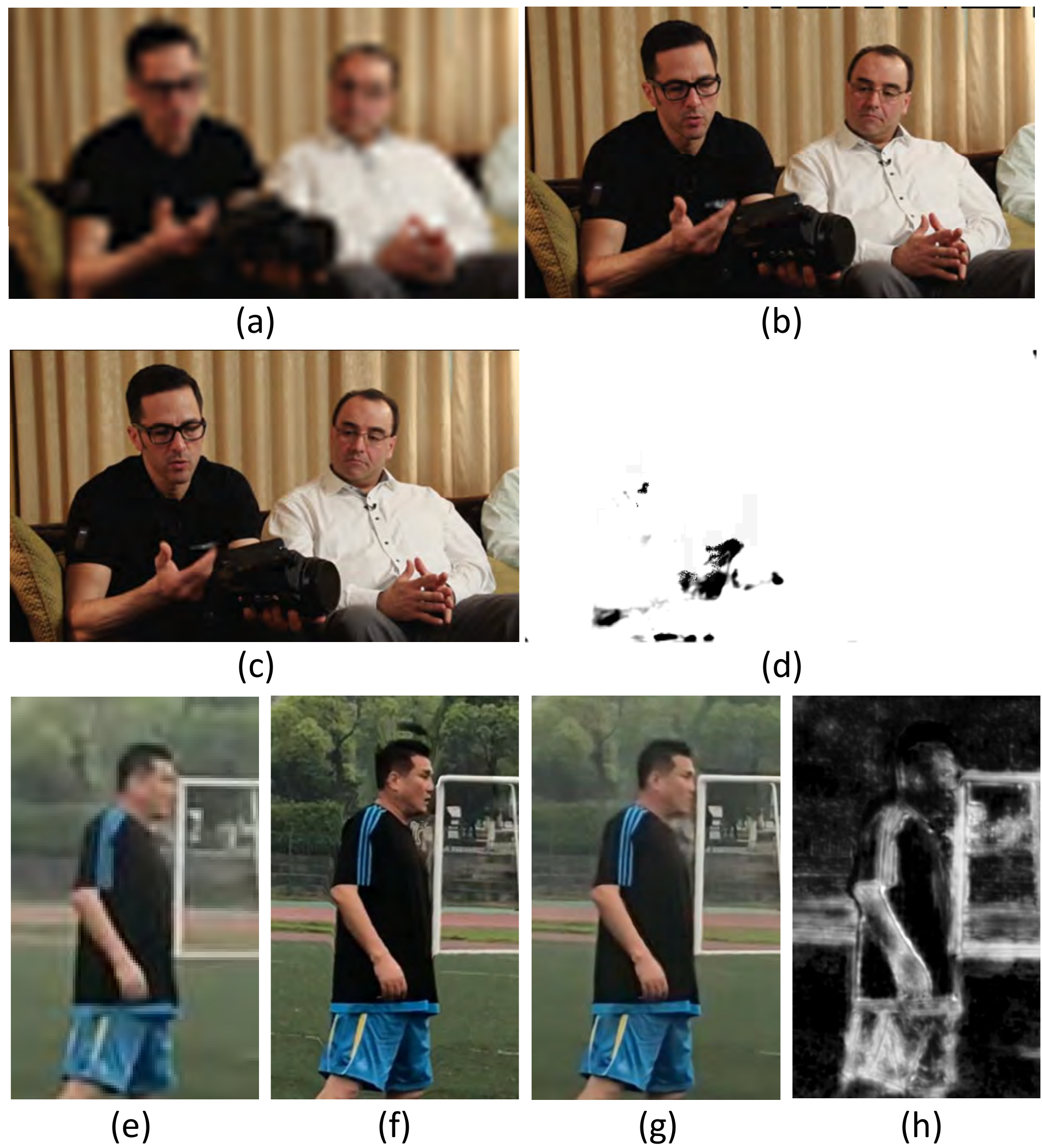}
    \caption{{\bf Synthesized Quality and Weighting Map $W$.} (a) to (d) are exemplified for Vimeo90K simulation data: (a) is the up-scaled  $I_{\tt LSR\uparrow}$ using bicubic method from  $I_{\tt LSR}$ by 8$\times$ for both spatial dimensions; (b) is the warped reference frame $I_{\tt ref}^w$; (c) is the output synthesized frame $Y$; (d) is the adaptive weighting map $W$ on (b); (e) to (h) are the visualization for camera captured real data with 3$\times$ resolution scaling from $I_{\tt LSR}$ to $I_{\tt LSR\uparrow}$: (e) is the up-scaled image $I_{\tt LSR\uparrow}$ from the captured LSR-HFR frame; (f) is the warped reference frame $I_{\tt ref}^w$; (g) is the output synthesized frame $Y$; (h) is the adaptive weighting map $W$ on (f).}
    \label{fig:visualization_images_real_sim_data2}
\end{figure}
We perform the real video data capture using dual iPhone 7 and Grasshopper3 cameras. One represents a consumer mobile camera used massively and the other one a camera commonly used for scientific or industrial imaging applications.

\begin{figure}[t]
  \centering
\subfigure[Y with $\sigma^2$ = 0]{\includegraphics[scale=0.04]{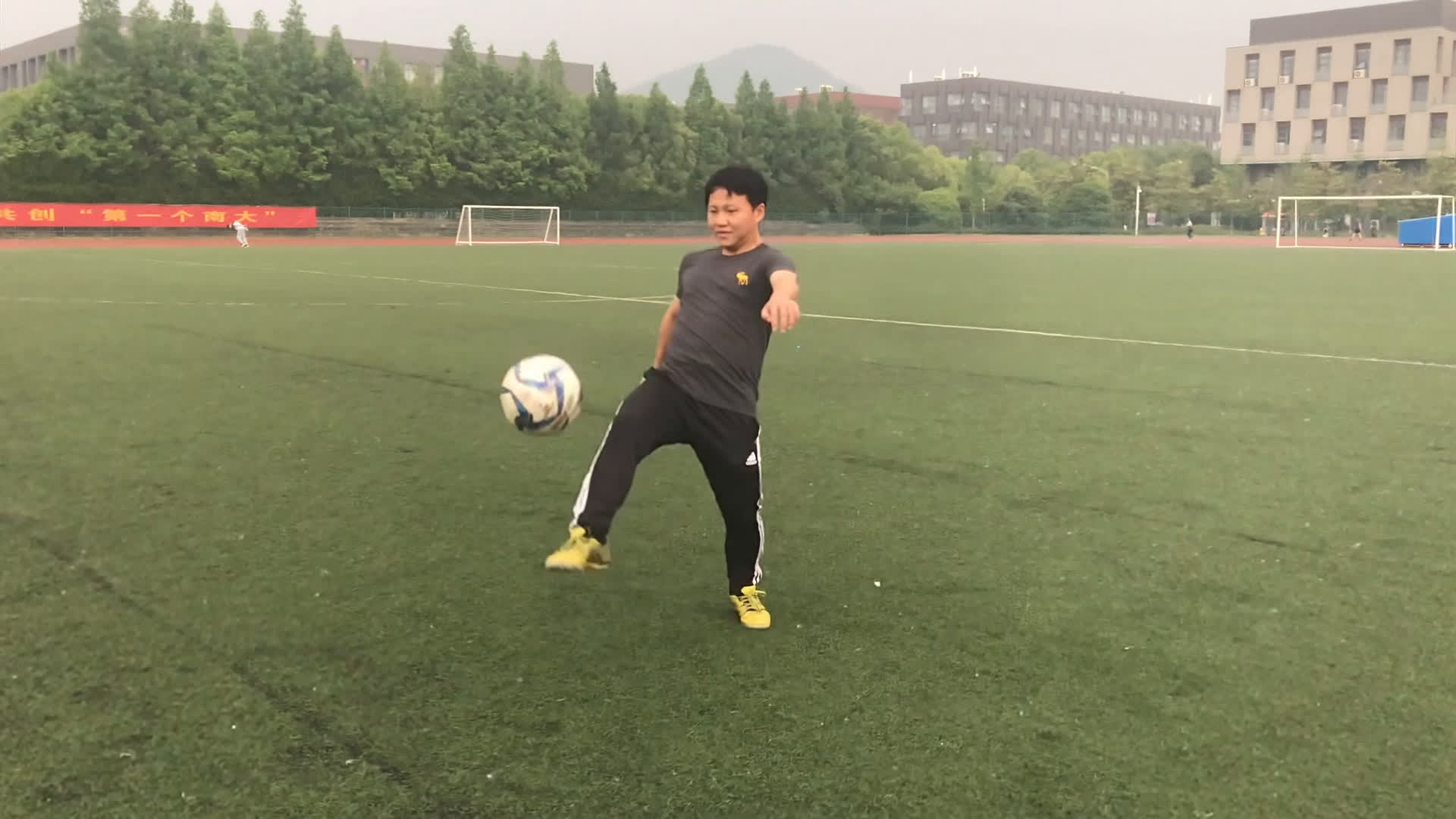}}
\subfigure[Y with $\sigma^2$ = 0.001]{\includegraphics[scale=0.04]{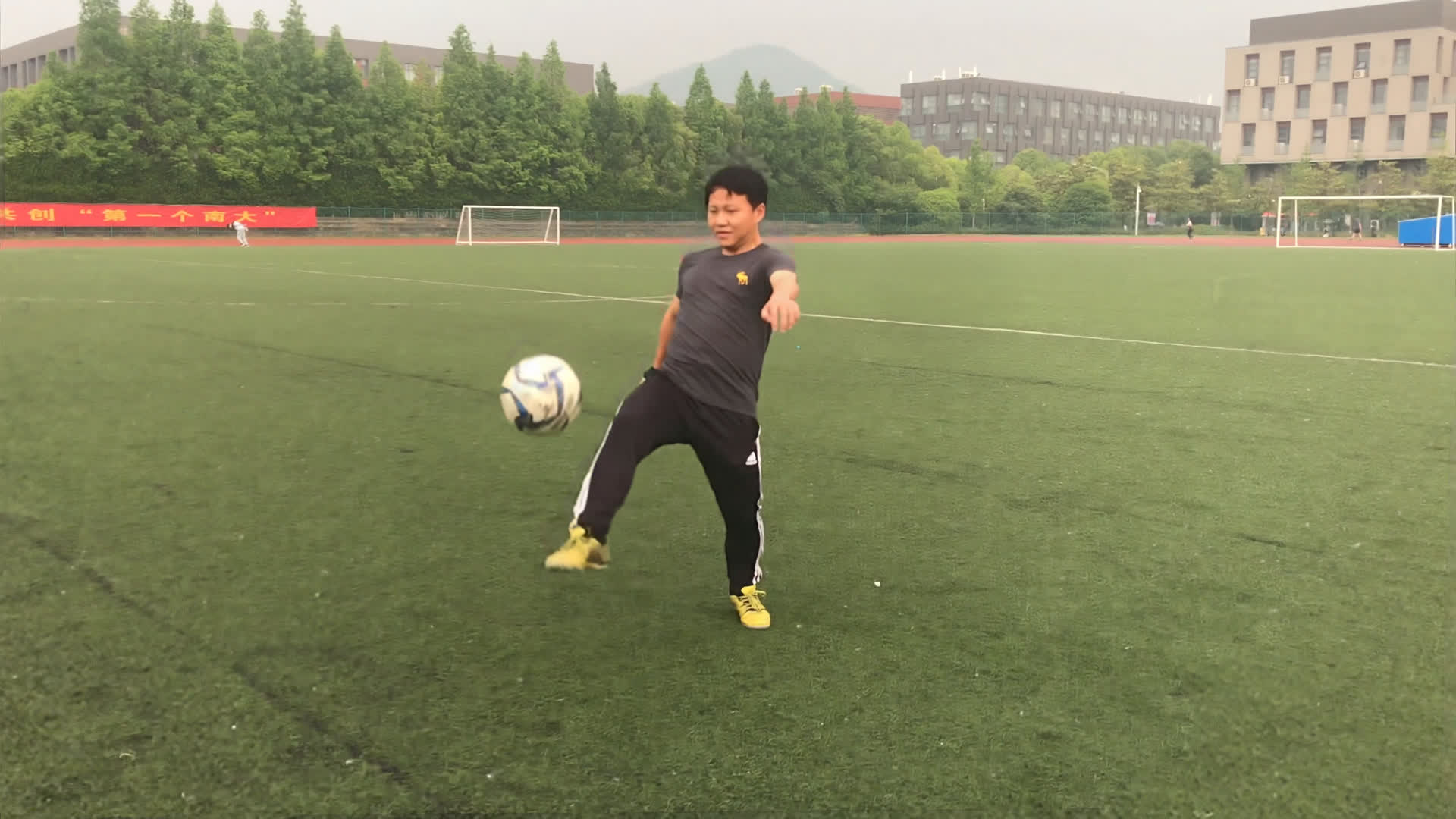}}
\subfigure[Y with $\sigma^2$ = 0.01]{\includegraphics[scale=0.04]{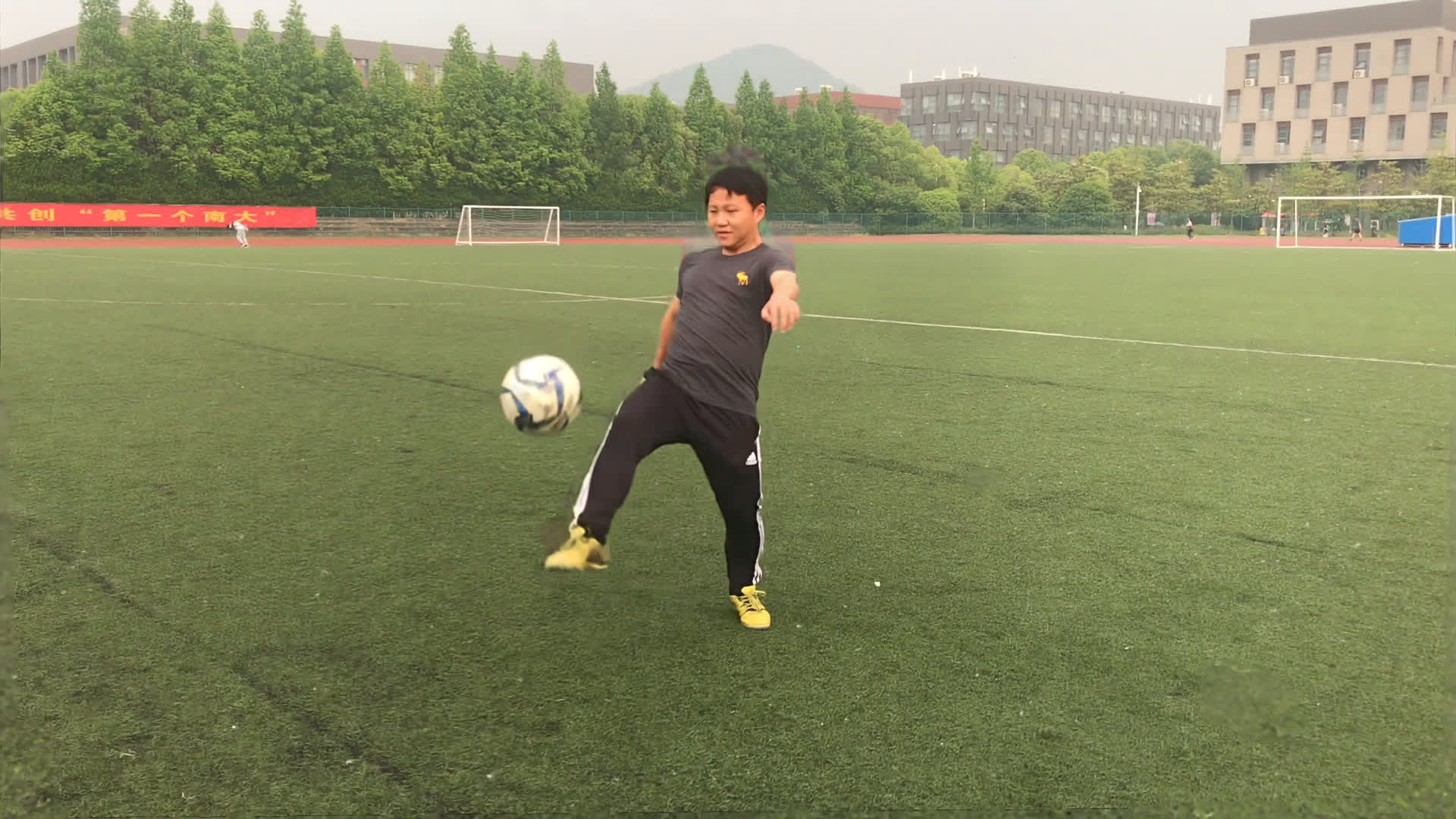}}\\
\subfigure[$W$ with $\sigma^2$ = 0]{\includegraphics[scale=0.04]{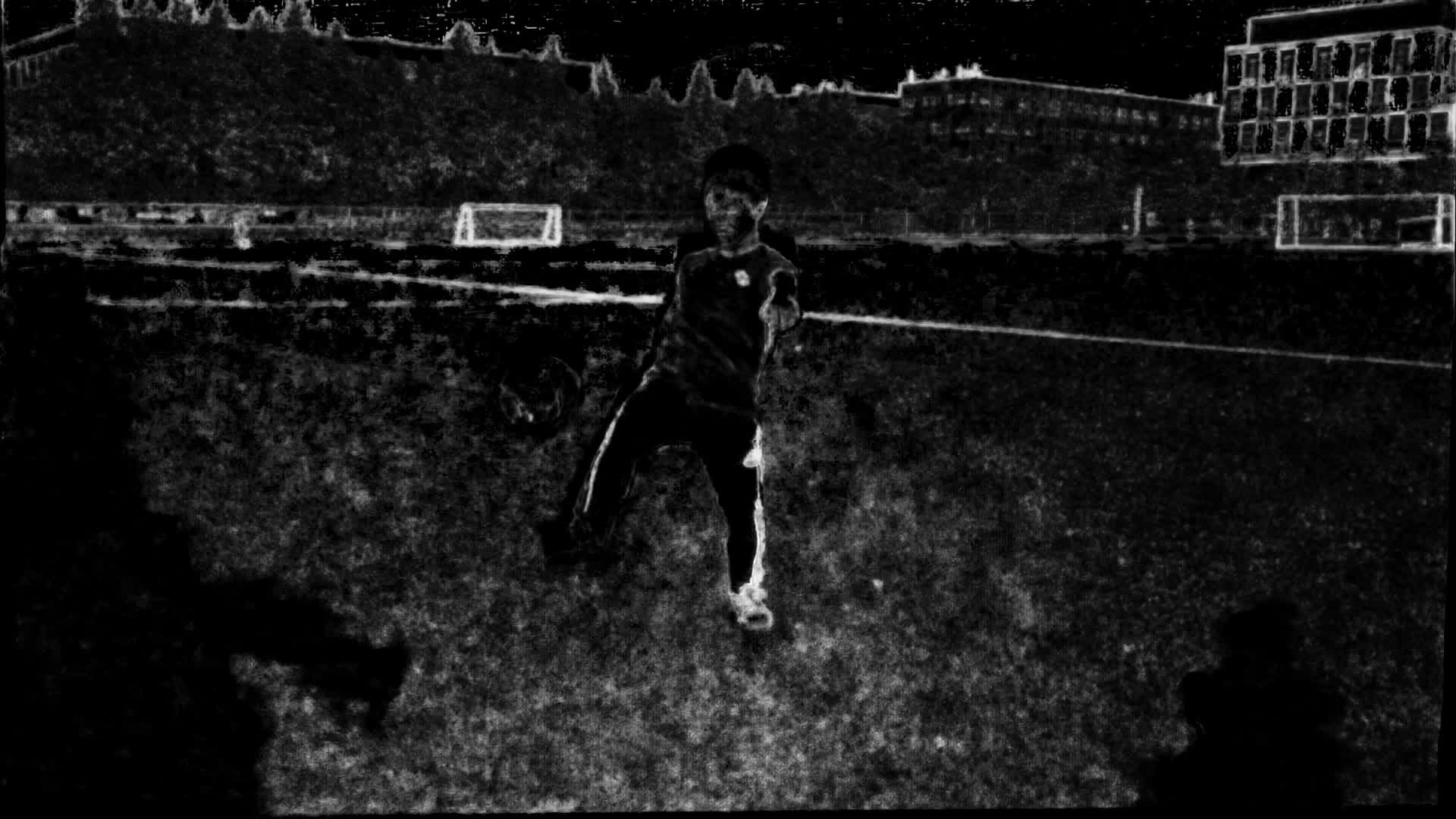}}
\subfigure[$W$ with $\sigma^2$ = 0.001]{\includegraphics[scale=0.04]{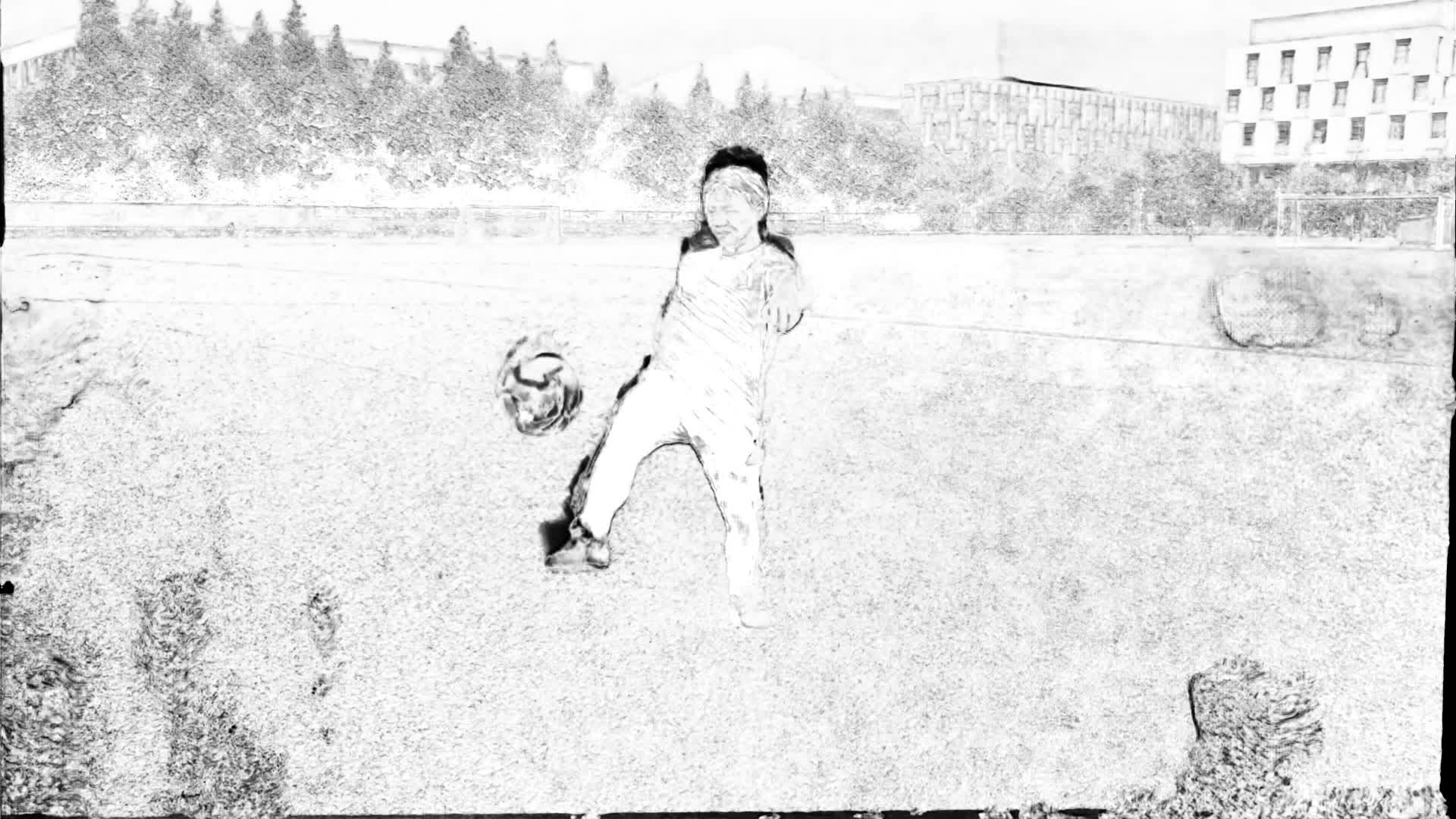}}
\subfigure[$W$ with $\sigma^2$ = 0.01]{\includegraphics[scale=0.04]{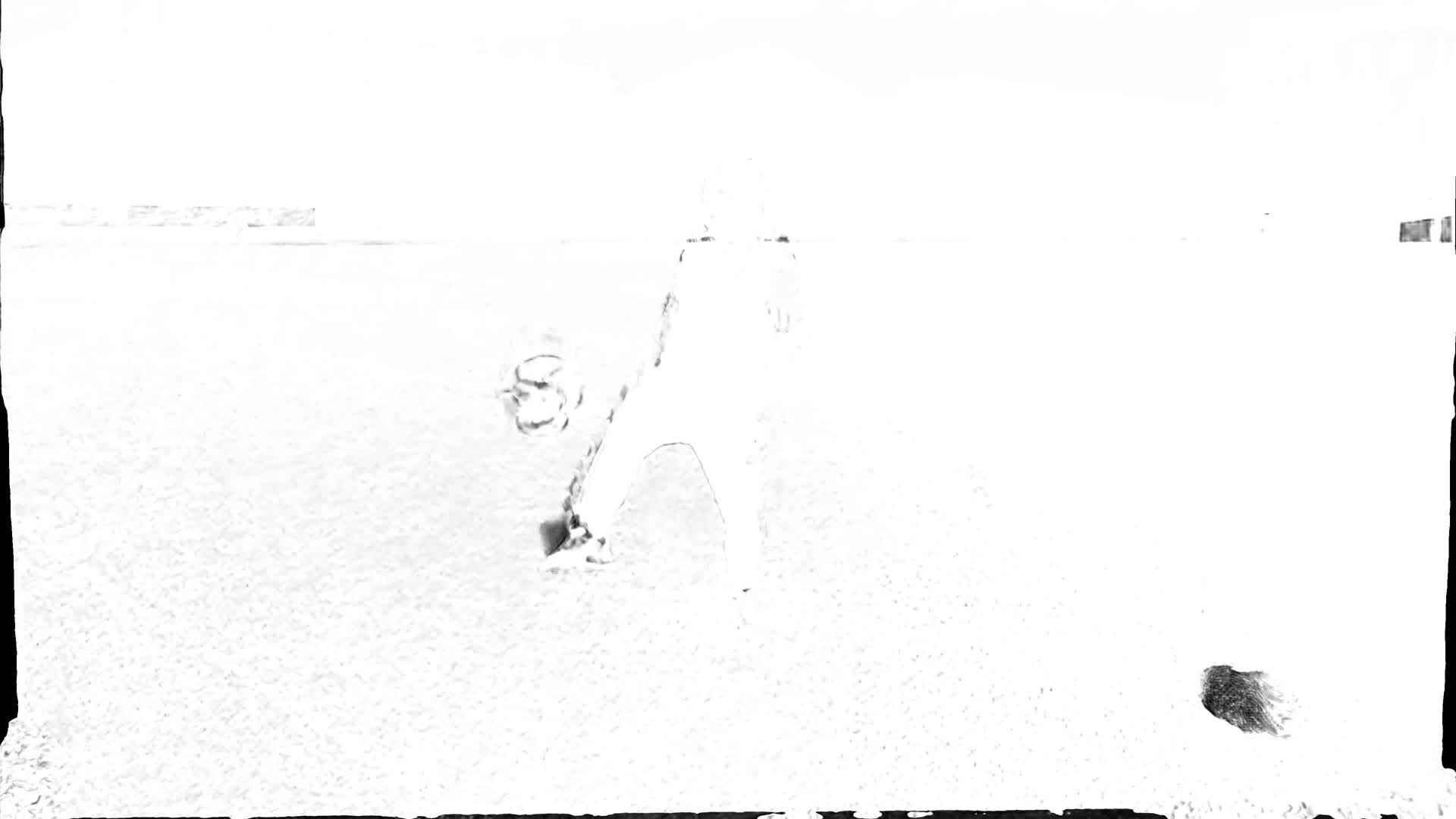}}
\caption{\textbf{Noise Regularization.} (a)-(c) Reconstructions of synthesized $Y$ with  noise level $\sigma^2$ at 0, 0.001 and 0.01; (d)-(f) Weighting map $W$ with noise level $\sigma^2$ at 0, 0.001 and 0.01. Video frames are captured using dual iPhone 7. Noise regularization shifts more weights to $I_{\tt ref}^w$ in general to improve the image quality, especially for those background stationary areas. But for those regions with occlusions (edge of the athletes), motion blurs (soccer ball) and warping artifacts (grassland), reconstruction still prefers pixels from $I_{\tt LSR\uparrow}$ to minimize the training loss. Ghosting artifact appear around the player's head in (b, c), which occur if the AWnet selects blurry regions from $I_{\tt LSR\uparrow}$ for the pixels that are not correctly aligned in $I_{\tt ref}^{w}$. }
  \label{fig:noise_regularization_impact}
\end{figure}

\subsubsection{Camera {Alignment}}
Dual camera setup requires careful calibration to map their relative coordinates and poses; especially, if we move the system around for shooting different scenes.
As suggested by~\cite{liu2013bundled}, we choose mesh-based homography for alignment. which greatly improves the accuracy of subsequent optical flow derivation as shown in Fig.~\ref{fig:mesh}.
More specifically, we extract SURF features~\cite{bay2008speeded} for both HSR-LFR and LSR-HFR frames, and then use the matched feature points to derive the homography transformation matrix for alignment. Fig.~\ref{fig:mesh}(c) shows that optical flow-based alignment fails to produce a good image. We then show the HSR-LFR frame aligned using the mesh-based homography in Fig.~\ref{fig:mesh}(d).  Fig.~\ref{fig:mesh}(e) reveals that homography-based alignment followed by optical flow-based refinement greatly improves the image quality.



In temporal dimension, for dual iPhone 7 configuration, we use a millisecond timer to synchronize two cameras. Synchronization with industrial cameras is much easier where we can apply the hardware clock timer.

Our dual camera system is connected to a high performance computer for both HSR-LFR and LSR-HFR video caching. The computer is using an Intel Xeon E5-2620 running at 2.10 GHz  with a GeForce GTX 1080Ti GPU. Without any platform and algorithmic optimization, it now takes about 7.4 seconds to synthesize a $3840\times2160$ frame from a pair of $1280\times720$ frames at 240FPS and $3840\times2160$ at 30FPS videos.

\subsubsection{Noise Regularization}
We first directly applied our model on data captured with real camera, but we  found that the quality of synthesized images was not as good as we had in the case of simulation data, as shown in Fig.~\ref{fig:visualization_images_real_sim_data2} (see (c) versus (g)). We observe that the reconstruction of data captured with iPhone 7 camera is blurry in  Fig.~\ref{fig:visualization_images_real_sim_data2}(g), even though the reference image after warping retains sharp details, as shown in Fig.~\ref{fig:visualization_images_real_sim_data2}(f).


Recall that the synthesized video frame $Y$ in \eqref{eq:output_Y} is a weighted combination of the warped reference $I_{\tt ref}^w$ and upscaled LSR-HFR input $I_{\tt LSR\uparrow}$. The relative weighting factor describing the weight contribution from the warped reference for a pixel at $(x,y)$-th position can be written as
\begin{align}\label{equ:weight}
W&(x,y) = \nonumber\\
&\frac{M(x,y)\sum_{i,j=1}^{5} K_{x,y} (i,j)}{(1-M(x,y))+M(x,y)\sum_{i,j=1}^{5} K_{x,y}(i,j)}.
\end{align}
Note that the value of $W(x,y)$ lies within [0, 1]. The larger the  $W(x,y)$ is, the more the $I_{\tt ref}^w$ contributes and vice versa.
As revealed in Fig.~\ref{fig:visualization_images_real_sim_data2}(e), and Fig.~\ref{fig:visualization_images_real_sim_data2}(f), our model shifts more weights to $I_{\tt LSR\uparrow}$, rather $I_{\tt ref}^w$ for captured real image synthesis, resulting in over smoothed reconstruction of $Y$.

We also observe that $I_{\tt ref}^w$ in Fig.~\ref{fig:visualization_images_real_sim_data2}(f) not only preserves the sharp details but also the rich colors compared to $Y$ in Fig.~\ref{fig:visualization_images_real_sim_data2}(g). One potential cause for this is that the camera parameters adjust automatically during recording to improve the image quality (e.g., ISO setting, aperture size, etc.). Therefore, the resulting LSR-HFR video has a narrow dynamic range and low SNR, as shown in Fig.~\ref{fig:visualization_images_real_sim_data2}(e) versus Fig.~\ref{fig:visualization_images_real_sim_data2}(f) from an associated HSR-LFR camera. In another words, the overall quality of $I_{\tt LSR}$ (or $I_{\tt LSR\uparrow}$) is much worse compared to the corresponding $I_{\tt ref}$ (or $I_{\tt ref}^w$).


In the case of simulation data, we do not face this problem because both $I_{\tt LSR}$ and $I_{\tt ref}$ are generated from the same ground truth. Such phenomena are also observed when using other RefSR methods (e.g., PM, CrossNet) to super-resolve  real data. The differences between the simulated data and the real sensor data affect the accuracy of optical flow and the performance of FusionNet. The FusionNet relies on the similarity between $I_{\tt LSR\uparrow}$ and $I_{\tt ref}^w$ to combine them. The higher resolution of real sensor data and the difference between camera parameters cause errors in optical flow estimation and mislead the FusionNet to pay less attention to $I_{\tt ref}^w$ as shown in Fig.~\ref{fig:visualization_images_real_sim_data2}(h).
Thus, it seems that models trained using ``clean'' simulation data can not be  directly extended to camera captured real data.

To apply our model on real data, we formulate a ``regularization'' problem that searches for a better weighting factor $W$ between $I_{\tt LSR\uparrow}$ and $I_{\tt ref}^w$ in \eqref{eq:output_Y}. We propose to add {noise} $n$ in $I_{\tt LSR\uparrow}$ for regularization during the  training progress. Thus, end-to-end learning optimization in \eqref{equ:l1loss} can be updated using regularized $Y$ and $I_{\tt ref}^w$. The only difference here is that instead of using noiseless $I_{\tt LSR\uparrow}$, we inject noise and use $I_{\tt LSR\uparrow}+n$ in all the computations.
With such noise regularization on $I_{\tt LSR\uparrow}$, network  learns to shift more weights to $I_{\tt ref}^w$ to improve the quality of synthesized reconstruction. Adding synthetic noise at the time of training provides robustness and acts as data augmentation \cite{zheng2016improving,noh2017regularizing,eilertsen2017hdr,eilertsen2019single,xie2019self,gu2019blind,jin2019learning,zhang2019deep}.

We train our network with Gaussian noise at different variances $\sigma^2$s, e.g., 0, 0.001, and 0.01. Snapshots of reconstructed $Y$ and weighting factor $W$ with various $\sigma^2$s are shown in Fig.~\ref{fig:noise_regularization_impact}. More comparisons can be seen in the supplementary material. We can see that reconstruction shifts more weights to $I_{\tt ref}^w$ (e.g., elements in $W$ gets closer to 1 in Fig.~\ref{fig:noise_regularization_impact}(d)--(f) as the noise increases (i.e., $\sigma^2$ increases), yielding better image quality with higher dynamic range, better color and sharper details (see Fig.~\ref{fig:noise_regularization_impact}(a)--(c)). Misalignment due to camera parallax and occlusions can introduce artifacts in the recovered images as can be seen around the player's head.
If we increase $\sigma^2$ further, the quality of images deteriorates. Therefore, to demonstrate the effect of noise regularization, we apply the noise regularization with $\sigma^2$ = 0.01 for all evaluations and comparisons in subsequent sections.

\begin{figure*}[htbp]
    \centering
    \includegraphics[width=0.9\textwidth]{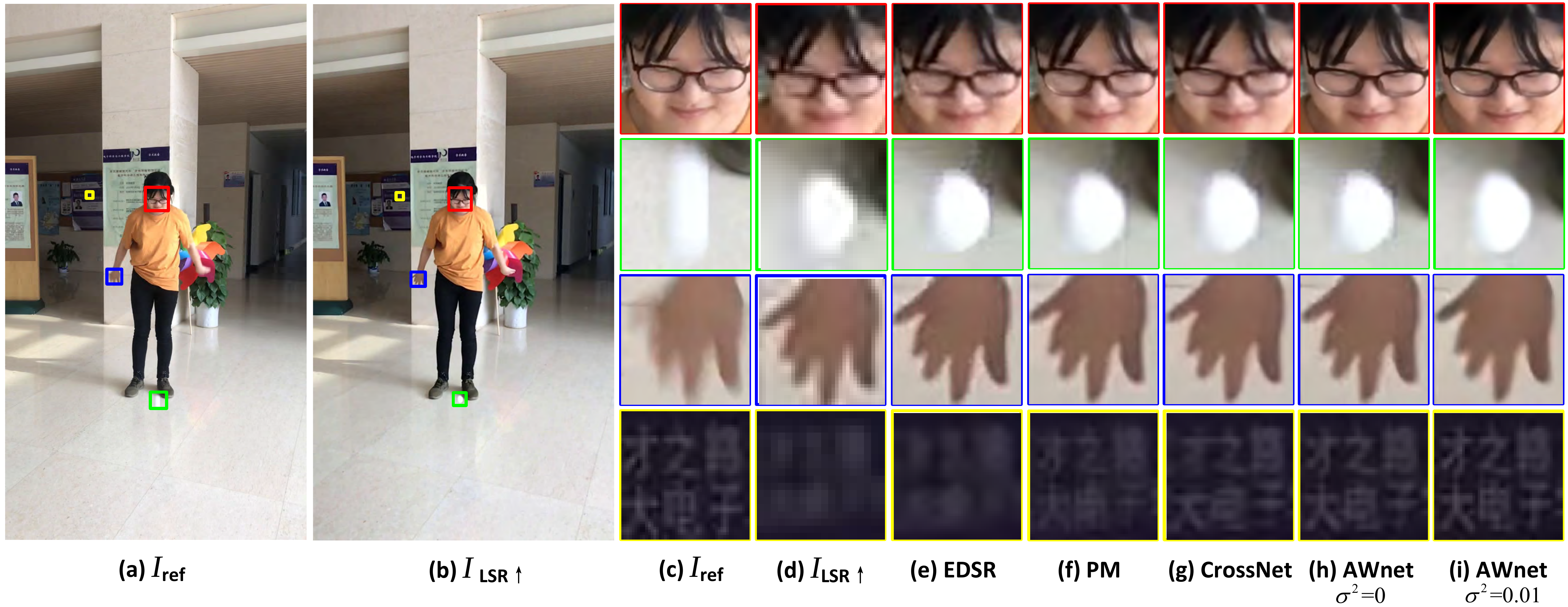}
    \caption{{\bf Super-Resolution:} Indoor activity with medium light illumination. $I_{\tt ref}$ is the captured 4k frame from the HSR-LFR camera, and synchronized $I_{\tt LSR\uparrow}$ is the up-scaled frame from the captured 720p frame of the LSR-HFR camera. Zoomed-regions are visualized for (c) $I_{\tt ref}$, (d) $I_{\tt LSR\uparrow}$, and  super-resolved reconstructions using (e) EDSR, (f) PM, (g) CrossNet, (h) AWnet with $\sigma^2 =0$ (no noise regularization), and (i) AWnet with $\sigma^2 =0.01$.}
    \label{fig:indoor_m_light}
\end{figure*}

\begin{figure*}[htbp]
    \centering
    \includegraphics[width=0.9\textwidth]{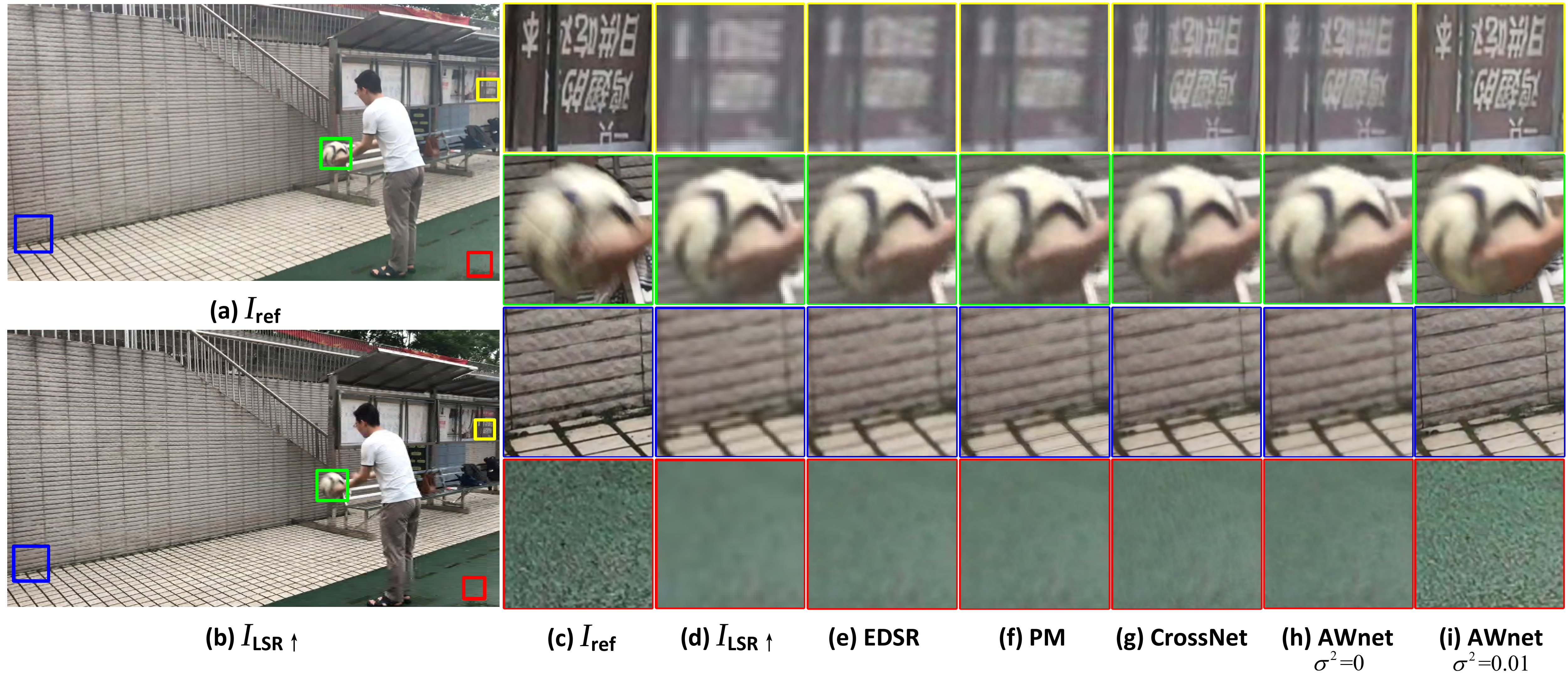}
    \caption{{\bf Super-Resolution:} Outdoor activity with low light illumination. $I_{\tt ref}$ is the captured 4k frame from the HSR-LFR camera, and synchronized $I_{\tt LSR\uparrow}$ is the up-scaled frame from the captured 720p frame of the LSR-HFR camera. Zoomed-regions are visualized for (c) $I_{\tt ref}$, (d) $I_{\tt LSR\uparrow}$, and  super-resolved reconstructions using (e) EDSR, (f) PM, (g) CrossNet, (h) AWnet with $\sigma^2 =0$ (no noise regularization), and (i) AWnet with $\sigma^2 =0.01$.} 
    \label{fig:outdoor_low_light}
\end{figure*}

\subsubsection{Subjective Evaluation}

For data captured with the real cameras, we compare the performance of our method with the EDSR~\cite{lim2017enhanced}, PM~\cite{boominathan2014improving}\footnote{Because the size of camera captured video frame is larger than the simulation content in Vimeo90K, we enlarge the patch size from 8 to 16 and search range from 16 to 64 for patch matching.}, and CrossNet~\cite{zheng2018crossnet}.

{\bf Super-Resolution.} Dual iPhone 7 cameras are used in this study. One camera captures a 4K video at 30FPS as the HSR-LFR input, and the other synchronized camera records 720p video at 240FPS as the LSR-HFR input.


%
We shoot videos for different scenes to validate the efficiency and generalization of our system. These scenes include indoor and outdoor activities with different illumination conditions, as illustrated in Figs.~\ref{fig:indoor_m_light}, and~\ref{fig:outdoor_low_light}. We can  observe the quality improvements of our proposed method when compared with the CrossNet~\cite{zheng2018crossnet}, PM~\cite{boominathan2014improving}, and EDSR~\cite{lim2017enhanced}. With appropriate noise regularization (e.g., $\sigma^2$ = 0.01 as exemplified), we could clearly observe that both the spatial details of $I_{\tt ref}$ and accurate motions from $I_{\tt LSR}$ are well retained and synthesized in the final reconstruction. We also notice that the subjective quality improvement is perceivable in our method with low and medium light illumination. With strong light illumination, the state-of-the-art CrossNet also provides good  reconstruction, but our method still provides the best results, as shown in supplemental material\footnote{The reconstructed videos are available  at \url{http://yun.nju.edu.cn/d/def5ea7074/?p=/Illumination&mode=lis}.}.

\begin{figure*}[htbp]
  \centering
  \includegraphics[width=0.85\textwidth]{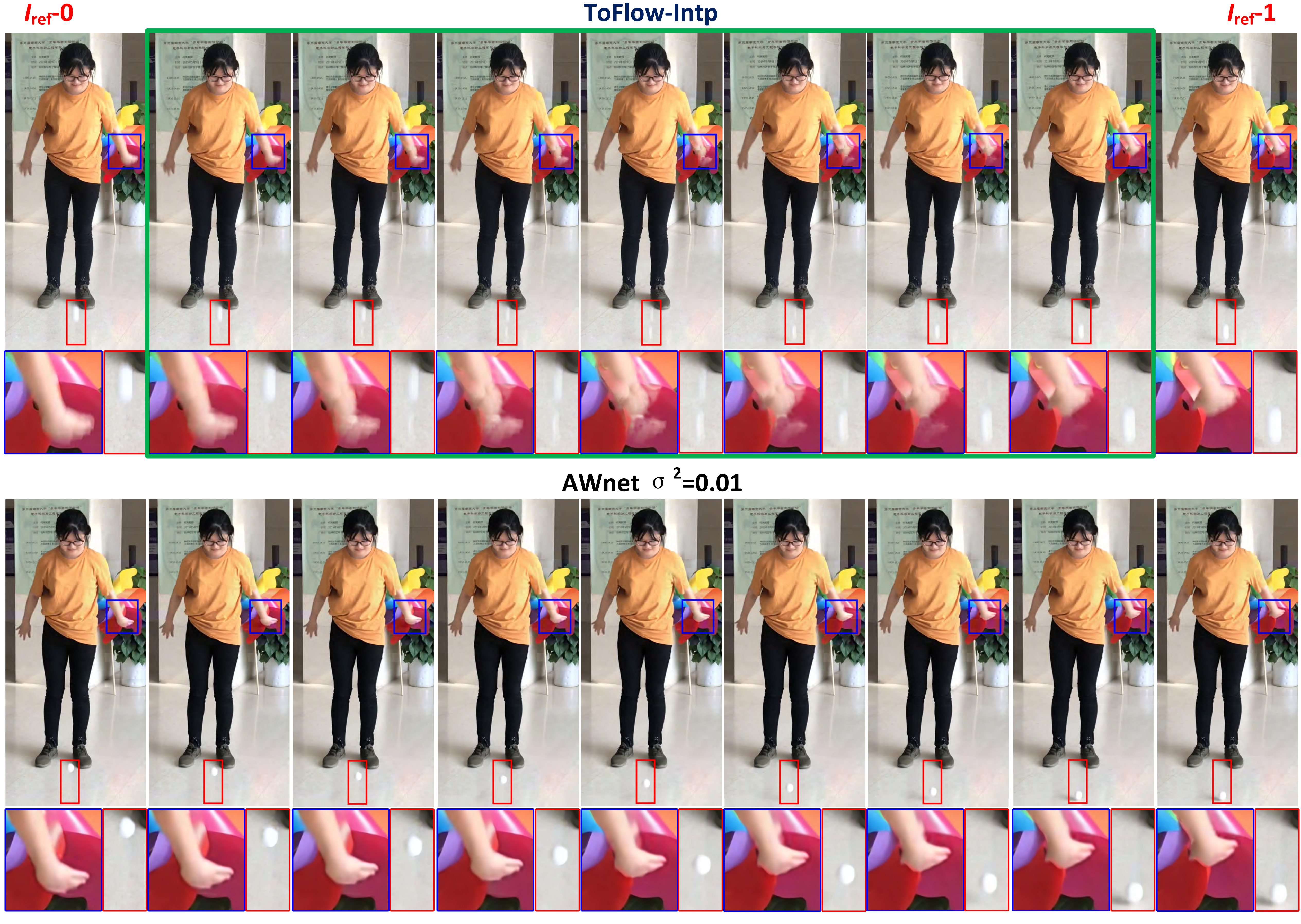}
  \caption{\textbf{Frame interpolation.} Indoor activity with medium light illumination. The most left and the most right of first rows are the captured HSR-LFR frames. Seven frames in-between are interpolated using ToFlow-Intp~\cite{xue2019video}; The second rows are the synthesized HSTR frames using our AWnet, which is trained with noise regularization with the variance of 0.01. \textbf{Zoom in the pictures, and you will see more image details.} (More in supplemental material.) }
  \label{fig:frame_interp_pingpong}
\end{figure*}

 \textbf{Frame Interpolation.} We extend our evaluations to frame interpolation. We present the entire GoP reconstructions in  Fig. \ref{fig:frame_interp_pingpong} for subjective comparison. $I_{\tt ref}$-0 and $I_{\tt ref}$-1 are the original frames from our HSR-LFR camera, while the frames in-between interpolated using ToFlow-Intp~\cite{xue2019video} are presented in the upper rows (highlighted with the green box). For comparison, reconstructed $Y$ frames using our model are placed in the bottom rows. As we can see, ToFlow-Intp shows ghosting and motion blurring (e.g., almost invisible fast-dropping ping-pong ball in the upper part of Fig.~\ref{fig:frame_interp_pingpong}). Our proposed method recovers the high-fidelity spatial details (see the woman's face and the texts on the wall) and accurate motions (see the fast-moving ping-pong ball and the woman's hands) at the same time.

\section{Ablation Studies} \label{sec:ablation_studies}
In this section we investigate different parameters of our system individually to understand the system capability and the source of efficiency.

    \begin{figure*}[t]
    \centering
    \includegraphics[width=1.0\textwidth]{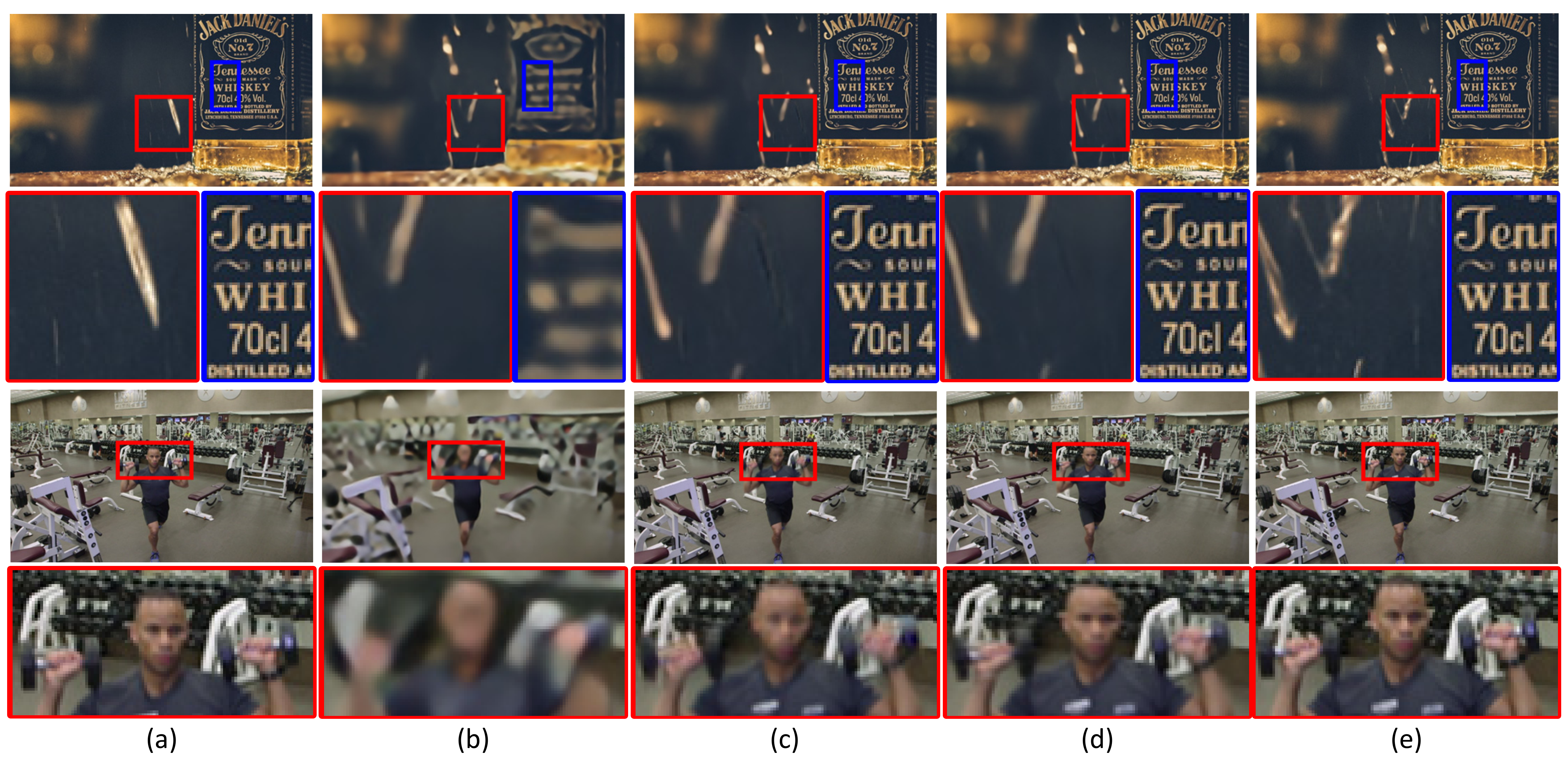}
    \caption{{\bf Subjective Evaluation of Adaptive Weighting Fusion  and Convolution-based Direction Prediction Using Synthetic Vimeo90K Test Dataset.} (a) $I_{\tt ref}$; (b) $I_{\tt LSR}$;  (c) Convolution-based Direct Prediction; (d) Adaptive Weighting Fusion; (e) Ground truth. The upper part is a raining scene with zoomed rain drop and label. The bottom part is a gym scene with zoomed face of a trainee.}
  	\label{fig:AW_vs_conv_vimeo}
    \end{figure*}	

\begin{table}[t]
\label{tab:sisr_impact}
\caption{Performance Impact of Different Upscaling Filter}
\centering
\renewcommand
\arraystretch{1.05}
\begin{tabular}{ c  c  c  c  c }
\specialrule{0em}{1pt}{1pt}
\hline
       \multirow{2}{*}{SISR } & \multicolumn{2}{c}{$4\times$} & \multicolumn{2}{c}{$8\times$}\\
\cline{2-5}
       & PSNR  & SSIM  & PSNR  &  SSIM  \\
\hline
      EDSR~\cite{lim2017enhanced} & 39.88 & 0.9862 & 36.63 & 0.9768\\
      bicubic & 39.75 & 0.9862 & 36.47 & 0.9766\\
\hline
\end{tabular}
\end{table}

{\bf Upscale Filter.} We upscale the $I_{\tt LSR}$ in Fig.~\ref{sfig:refSR} to the same resolution as the $I_{\tt ref}$ for subsequent processing. Previous explorations assume the state-of-the-art SISR method EDSR~\cite{lim2017enhanced}.
Here, we replace it with a straightforward bicubic filter. Models are re-trained with this new upscaling filter, and performance comparison is evaluated on the efficiency of respective $4\times$ and $8\times$ super-resolution application using the Vimeo90K testing dataset. Averaged PSNR and SSIM are listed in Table~\ref{tab:sisr_impact}, showing that different upscaling method does not affect the overall performance noticeably, e.g., $\approx$ 0.1 dB PSNR  and  $<=$ 0.002 SSIM index variations reported. This observation suggests that we can use simple upsampling filters to scale up $I_{\tt LSR}$ instead of complex super-resolution methods. In principle, this is mainly due to the fact that our   AWnet-based dual camera system could learn and embed high frequency spatial information from its HSR-LFR input for final reconstruction synthesis. Thus, complex super-resolution method used to estimate high frequency component is not an inevitable step any more.

\begin{table}[t]
	\caption{Objective PSNR of Reconstructed Images for {\it Single-Reference} and {\it Multi-Reference} AWnet.}
	\label{table:reference-x}
	\centering
	\renewcommand
	\arraystretch{1.05}
	\begin{tabular}{ c | c | c  }
		\specialrule{0em}{1pt}{1pt}
		\hline
		&  Single-Reference    &     Multi-Reference        \\ 
		\hline
		averaged & 35.33 dB     &   36.21  dB        \\  
		\hline
	\end{tabular}
\end{table}

{\bf Reference Structure.} We also compare the performance of {\it Single-Reference} and {\it Multi-Reference}  AWnet.
	We evaluate the {\it Single-Reference} and {\it Multi-Reference} AWnet with Vimeo90K test dataset (e.g. 7824 video sequences in total). For {\it Multi-Reference} AWnet, we use 7 frames in each video sequence; treat the first  and the last frame as the references; and recover the remaining five frames that are down-scaled $8\times$ along both horizontal and vertical directions. In comparison, {\it Single-Reference} AWnet is configured same as described in Section~\ref{sec:perf_comp_sim_data}. The averaged PSNRs for all test videos are shown in Table~\ref{table:reference-x}. We observe that {\it Multi-reference} AWnet offers 0.9 dB gain compared to {\it Single-reference}  AWnet. Subjective quality is also improved with better temporal continuity and spatial texture details by introducing the multiple reference in AWnet\footnote{The videos for multi-reference AWnet results are available at  \url{http://yun.nju.edu.cn/d/def5ea7074/?p=/MultiReferenceAWnet&mode=list}}.
	Since CrossNet uses only one reference frame, we only present results for {\it Single-Reference} structure in other comparative studies.
	

{\bf FusionNet.} In our proposed method, FusionNet uses adaptive weighting fusion (AWFusion) on upscaled and warped images to produce output pixels. Alternatively, we can also apply convolution-based direct prediction. For such convolution-based direct prediction (ConvDP), we replace the U-net style architecture for dynamic filter and mask generation in Fig.~\ref{sfig:FusionNet} with stacked convolutions. Here, we utilize 36 convolutional layers with the same $3\times 3$ kernel, and nonlinear ReLU activation. The remaining parts of the AWnet framework are kept without change. We then evaluate the default AWFusion and ConvDP using the Vimeo90K test dataset (e.g., 7824 video sequences in total). Table~\ref{table:conv} provides the averaged PSNR and SSIM results of the reconstructions using respective methods. Although the quantitative gains in terms of PSNR and SSIM are not very large, Fig.~\ref{fig:AW_vs_conv_vimeo} shows noticeable quality improvement of synthesized frame with less noise and sharp/clear reconstruction for the AWFusion compared to ConvDP (e.g., Fig.~\ref{fig:AW_vs_conv_vimeo}(c) vs (d)).

\begin{table}[t]
	\caption{Objective Comparison between Adaptive Weighting Fusion and Convolution-based Direct Prediction for Output Pixel Reconstruction.}
	\label{table:conv}
	\centering
	\begin{tabular}{ c  c  c  c  c }
		\specialrule{0em}{1pt}{1pt}
		\hline
		& \multicolumn{2}{c}{4$\times$} & \multicolumn{2}{c}{8$\times$} \\
		\cline{2-5}
		&    PSNR & SSIM & PSNR & SSIM  \\
		\hline
		AWFusion &  39.88   &  0.9862  &  36.63   &  0.9768  \\
		ConvDP  &  39.67    & 0.9856  & 36.45   &  0.9756 \\
		\hline
	\end{tabular}
\end{table}

\textbf{Camera Parallax.} Dual camera setup is used in our system. Thus camera parallax could be an issue that affects the system performance. We show in our studies below by using simulation data from available KITTI~\cite{menze2015object}, Flower~\cite{srinivasan2017learning}, LFVideo~\cite{wang2017light}, Stanford light field~\cite{StanfordLF} datasets, and real data captured by our dual camera system with various parallax settings to demonstrate the robustness of our method.

\begin{table*}[htbp]
\caption{Objective Performance Comparison of 4$\times$ and 8$\times$ Super-Resolution Methods on Flower and LFVideo Datasets}
\label{table:flower_LFV}
\centering
\renewcommand
\arraystretch{1.05}
\begin{tabular}{ c  c  c  c  c c c c c c}
\specialrule{0em}{1pt}{1pt}
\hline
       \multirow{2}{*}{Methods} & \multirow{2}{*}{Scale} & \multicolumn{2}{c}{Flower(1,1)} & \multicolumn{2}{c}{Flower(7,7)} & \multicolumn{2}{c}{LFVideo(1,1)} & \multicolumn{2}{c}{LFVideo(7,7)} \\
\cline{3-10}
      &&PSNR & SSIM & PSNR & SSIM & PSNR & SSIM & PSNR & SSIM \\
\hline
      SRCNN~\cite{dong2015image} & $4\times$ & 32.76 & 0.89 & 32.96 & 0.90 & 32.98 & 0.86 & 33.27 &  0.86 \\
      VDSR~\cite{kim2016accurate} & $4\times$ & 33.34 & 0.90 & 33.58 & 0.91 & 33.58 & 0.87 & 33.87 &  0.88 \\
      MDSR~\cite{lim2017enhanced} & $4\times$ & 34.40 & 0.92 & 34.65 & 0.92 & 34.62 & 0.89 & 34.91 &  0.90 \\
      PM~\cite{boominathan2014improving} & $4\times$ & 38.03 & \textbf{0.97} & 35.23 & 0.94 & 38.22 & 0.95 & 37.08 & 0.94 \\
      CrossNet~\cite{zheng2018crossnet} & $4\times$ & 41.23 & 0.9625 & 38.09 & 0.9475 & 41.57 & \textbf{0.9758} & 39.17 & 0.9627\\
      {\bf AWnet} & $4\times$ & \textbf{41.33} & 0.9631 & \textbf{38.31} & \textbf{0.9492} & \textbf{41.63} & 0.9757 & \textbf{39.36} & \textbf{0.9635} \\
\hline
      SRCNN~\cite{dong2015image} & $8\times$ & 28.17 & 0.77 & 28.25 & 0.77 & 29.43 & 0.75 & 29.63 & 0.76 \\
      VDSR~\cite{kim2016accurate} & $8\times$ & 28.58 & 0.78 & 28.68 & 0.78 & 29.83 & 0.77 & 30.04 & 0.77 \\
      MDSR~\cite{lim2017enhanced} & $8\times$ & 29.15 & 0.79 & 29.26 & 0.80 & 30.43 & 0.78 & 30.65 & 0.79 \\
      PM~\cite{boominathan2014improving} & $8\times$ & 35.26 & 0.95 & 30.41 & 0.85 & 36.72 & 0.94 & 34.48 & 0.91 \\
      CrossNet~\cite{zheng2018crossnet} & $8\times$ & \textbf{39.35} & \textbf{0.9571} & 34.11 & 0.9149 & \textbf{40.63} & \textbf{0.9727} & 36.97 & 0.9465\\
      {\bf AWnet} & $8\times$ & 39.29 & \textbf{0.9571} & \textbf{34.53} & \textbf{0.9199} & 40.48 & 0.9725 & \textbf{37.25} & \textbf{0.9487} \\
\hline
\end{tabular}
\end{table*}

\begin{table}[htbp]
\caption{Objective Performance (PSNR) Comparison of 8$\times$ Super-Resolution Methods on Stanford Light Field Dataset}
\label{table:stanford}
\centering
\renewcommand
\arraystretch{1.05}
\begin{tabular}{ c  c  c  c  c}
\specialrule{0em}{1pt}{1pt}
\hline
Methods & parallax = (1,0) & (3,0) & (5,0) \\
\hline
MDSR~\cite{lim2017enhanced} & 29.66 & 29.66 & 29.67 \\
PM~\cite{boominathan2014improving} & 34.61 & 32.55 & 30.42 \\
CrossNet~\cite{zheng2018crossnet} & 39.33 & 36.77 & 35.15 \\
{\bf AWnet} & \textbf{39.53} & \textbf{37.65} & \textbf{36.47} \\
\hline
\end{tabular}
\end{table}

\begin{table}[htbp]
\caption{Performance Evaluation Using KITTI dataset for Super-Resolution}
\label{table:kitti}
\centering
\renewcommand
\arraystretch{1.05}
\begin{tabular}{ c  c  c  c  c }
\specialrule{0em}{1pt}{1pt}
\hline
       \multirow{2}{*}{} & \multicolumn{2}{c}{4$\times$} & \multicolumn{2}{c}{8$\times$}\\
\cline{2-5}
       & PSNR  & SSIM  & PSNR  &  SSIM  \\
\hline
      EDSR~\cite{lim2017enhanced} & 27.03 & 0.8519 & 23.47 & 0.7377\\
      CrossNet~\cite{zheng2018crossnet} & 27.43 & 0.8631 & 24.92 & 0.7981\\
      \textbf{AWnet} & \textbf{28.19} & \textbf{0.8882} & \textbf{26.01} & \textbf{0.8356}\\
\hline
\end{tabular}
\end{table}

{\it Comparison Using Simulation Data:} We test our AWnet on Flower~\cite{srinivasan2017learning}, LFVideo~\cite{wang2017light} and Stanford light field (Lego Gantry)~\cite{StanfordLF} datasets following the same configuration in~\cite{zheng2018crossnet}. The Flower and LFVideo datasets are light field images captured using Lytro ILLUM camera. {{Each light field image has 376$\times$541 spatial samples and 14$\times$14 angular samples (grid). The same as the methods applied in~\cite{srinivasan2017learning,zheng2018crossnet}, we
extract the central 8$\times$8 grid of angular samples to avoid invalid images.}} Parallax is offered by setting the reference image $I_{\tt ref}$ at (0, 0), and associated low-resolution
correspondence at ($i$, $i$), with $0< i\leq7$ by shifting position to another different angular sample. For example, Flower(1,1) and LFVideo(7,7) in Table~\ref{table:flower_LFV}, represent low-resolutions at (1,1) and (7,7) with respect to the respective references at (0,0). Images in both Flower and LFVideo datasets exhibit small parallax settings~\cite{zheng2018crossnet, srinivasan2017learning}.
On the other hand, Stanford light field dataset contains the light field images {{ shot using a Canon Digital Rebel XTi with a canon 10-22 mm lens. It is placed using a movable Mindstorms motor on the Lego gantry, where the parallax is introduced by the baseline distances along with the camera movement. Under such equipment settings, the captured light-field images have much larger parallax than those captured by Lytro ILLUM camera.}} Both Table~\ref{table:flower_LFV} and Table~\ref{table:stanford}
also shows the leading performance of our proposed AWnet at a variety of parallax between testing and reference images, further demonstrating the generalization of our network in different application scenarios. Especially, on average, up to 1.3 dB PSNR improvement is obtained of our AWnet against the CrossNet in Table~\ref{table:stanford} for large parallax setting. This is mainly due to the reason that CrossNet was not initially designed for RefSR with larger parallax.
Thus, additional parallax augmentation procedure was suggested in~\cite{zheng2018crossnet} for re-training.

KITTI dataset has 54cm baseline distance for two cameras. We apply our method and CrossNet with pretrained models using Vimeo90K dataset directly to KITTI test data with 400 stereo image pairs. We use the high-resolution left-view images as the reference for the low-resolution right-view images. Since both CrossNet and our method expect the image resolution to be divisible by 64, thus, we crop images to $1216\times 320$ for testing. Table~\ref{table:kitti} gives the PSNR and SSIM for super-resolution evaluation. Our method still offers better PSNR (e.g., $> 1$ dB gain for $8\times$ resolution scaling factor) and SSIM compared to the CrossNet. EDSR results are offered as a reference point, revealing that RefSR still exhibits superior performance, even with a large camera parallax (i.e., 54cm baseline in KITTI data). Results in Table~\ref{table:kitti} and Table~\ref{table:objective_perf_sim} suggest that both PSNR and SSIM are dropped significantly when evaluating models on KITTI compared to the Vimeo90K test data. This is because the introduction of the (large) camera parallax leads to the inaccurate flow estimation for later processing. Similar observations are reported in CrossNet~\cite{zheng2018crossnet} where PSNR and SSIM drop as the camera parallax increases.

\begin{figure*}[htbp]
  \centering
  \includegraphics[width=0.95\textwidth]{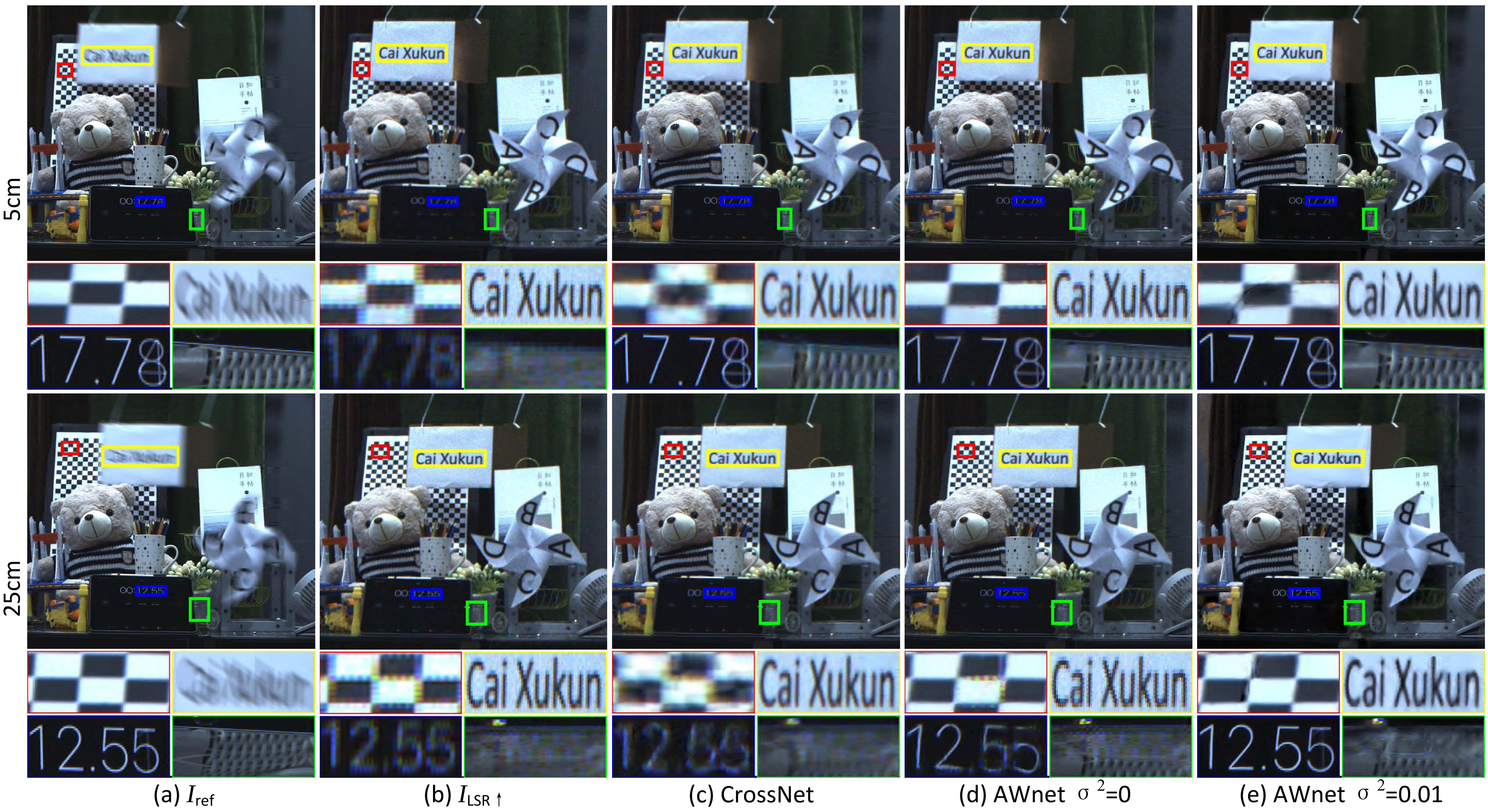}
  \caption{\textbf{Camera Parallax.} Image reconstruction for our dual camera system when placing cameras with baseline distance at 5cm and 25cm. Our LSR-HFR camera operates at 240FPS. These images are captured using dual Grasshopper3 GS3-U3-51S5C cameras. The frames in the first column are the captured HSR-LFR frames. The frames in the second column are the captured LSR-HFR frames. Look at the repeated patterns on the checkerboard snapshots, there are some ghosts on the results of CrossNet because its multi-scale warping in feature domain, but our method does not have this problem. And our method has strong robustness when parallax is large. \textbf{Zoom in the pictures, you will see more  details in the larger images.} More parallax settings in supplemental material.}
  \label{fig:camera_parallax}
\end{figure*}

\begin{figure*}[htbp]
  \centering
  \includegraphics[width=0.9\textwidth]{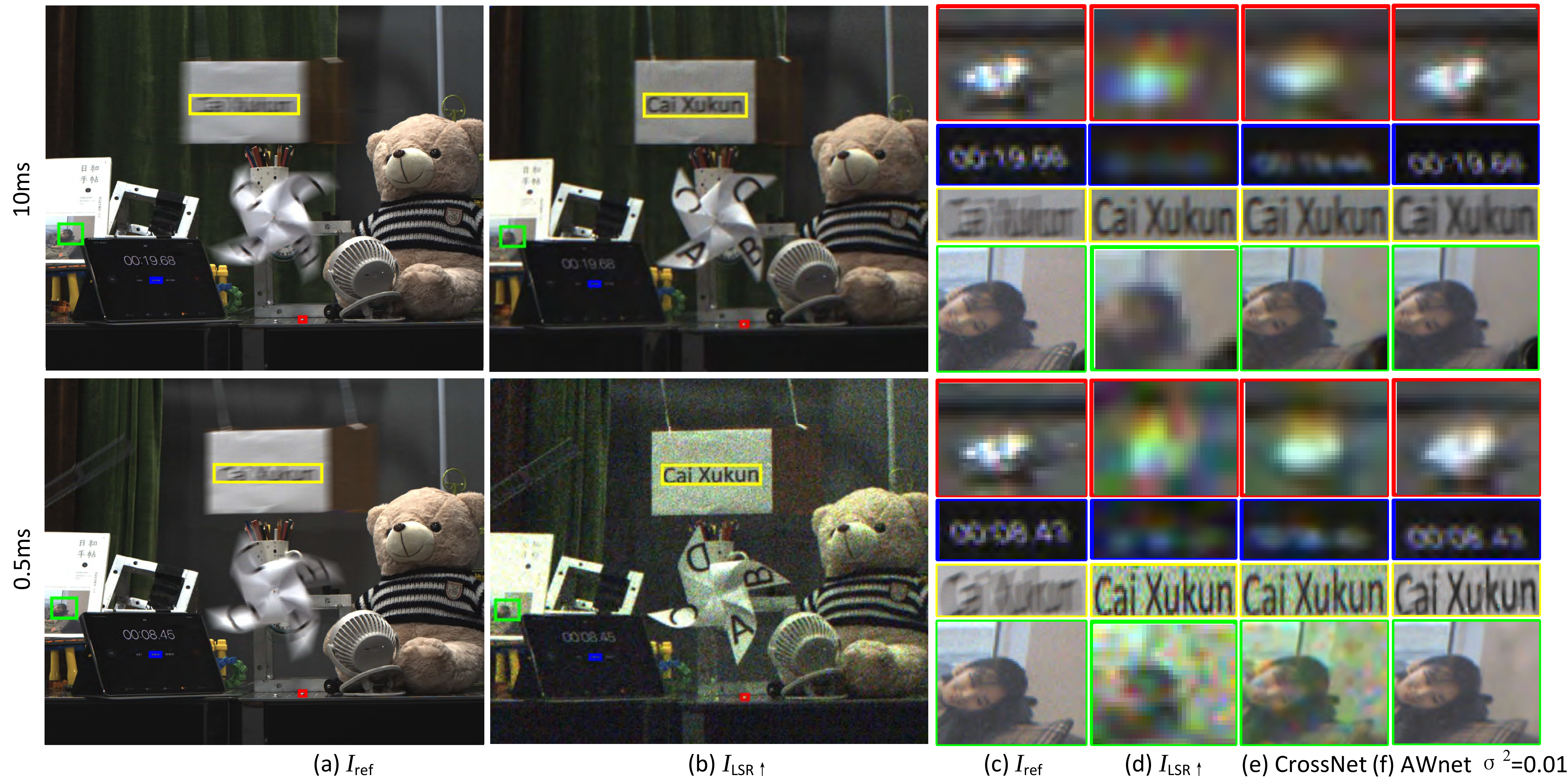}
  \caption{\textbf{Exposure Time.} Various exposure time  exemplified using our dual camera system. These images are captured using dual Grasshopper3 GS3-U3-51S5C cameras. The frames in the first column are the captured HSR-LFR frames using default exposure, the frames in the second column are the captured LSR-HFR frames with various exposure adjustments. Noise increases as exposure time decreases. CrossNet could remove some noise but generally lead to blurred artifacts induced by the global convolutions. Our AWnet can effective remove noise and improve the picture quality greatly. More exposure time settings in supplemental material.}
  \label{fig:exposure_noise2}
\end{figure*}

    \begin{figure}[t]
    \centering
    \includegraphics[width=0.5\textwidth]{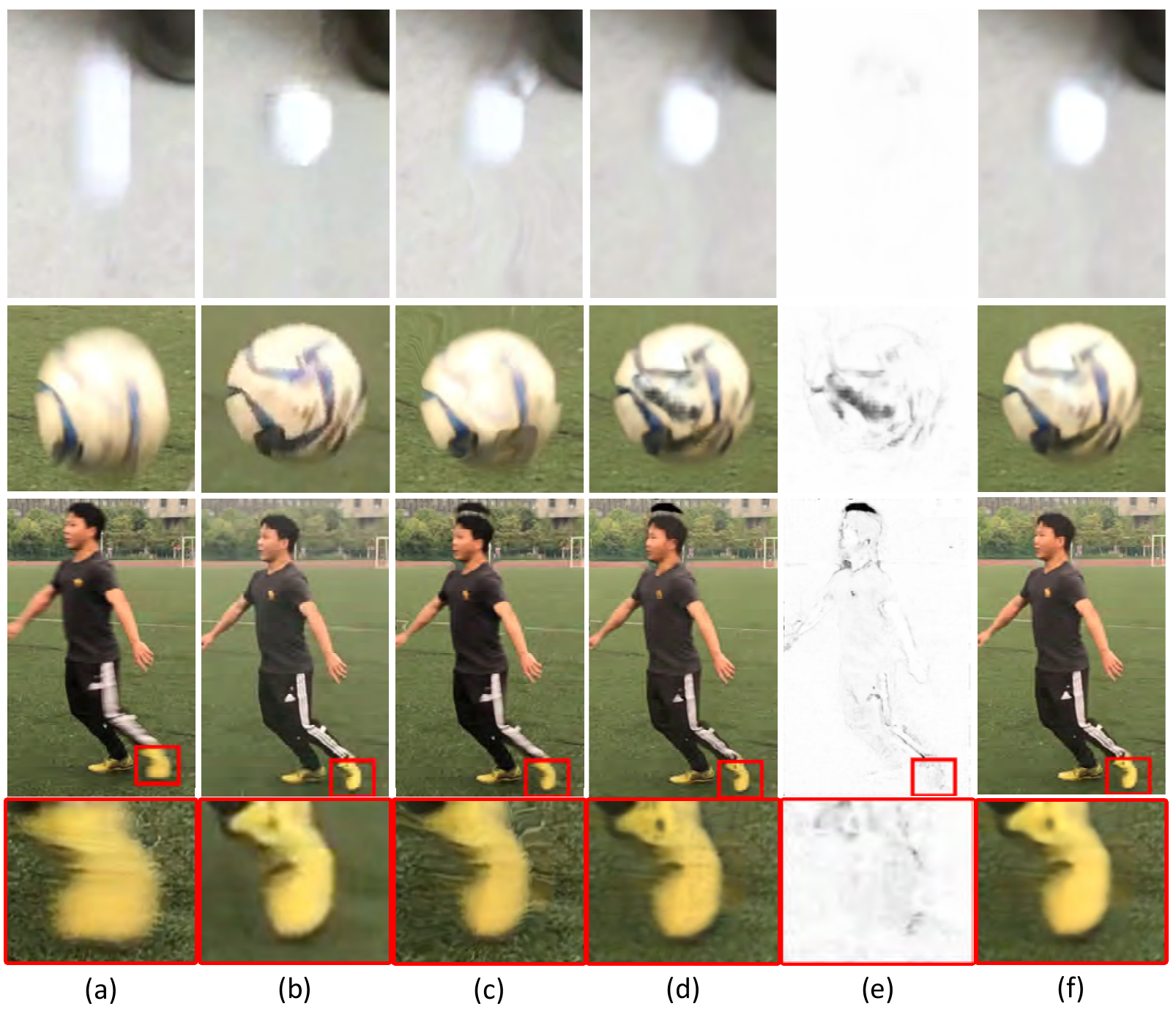}
    \caption{{\bf Motion blur in real data.} The images are captured with iPhone 7. (a) $I_{\tt HSR}$; (b) $I_{\tt LSR\uparrow}$; (c) $I_{\tt ref}^w$; (d) $I_{\tt ref}^{wk}$; (e) $M$; (f) $Y$ with $\sigma^2 = 0.01$. The first row shows a fast ping-pong. The second row shows a fast soccer ball. The third and fourth rows are a fast-moving player and his enlarged shoe. {\bf Zoom in to see more details.}}
  	\label{fig:motion_blur_new}
    \end{figure}

\begin{figure*}[htbp]
    \centering
    \includegraphics[width=0.85\textwidth]{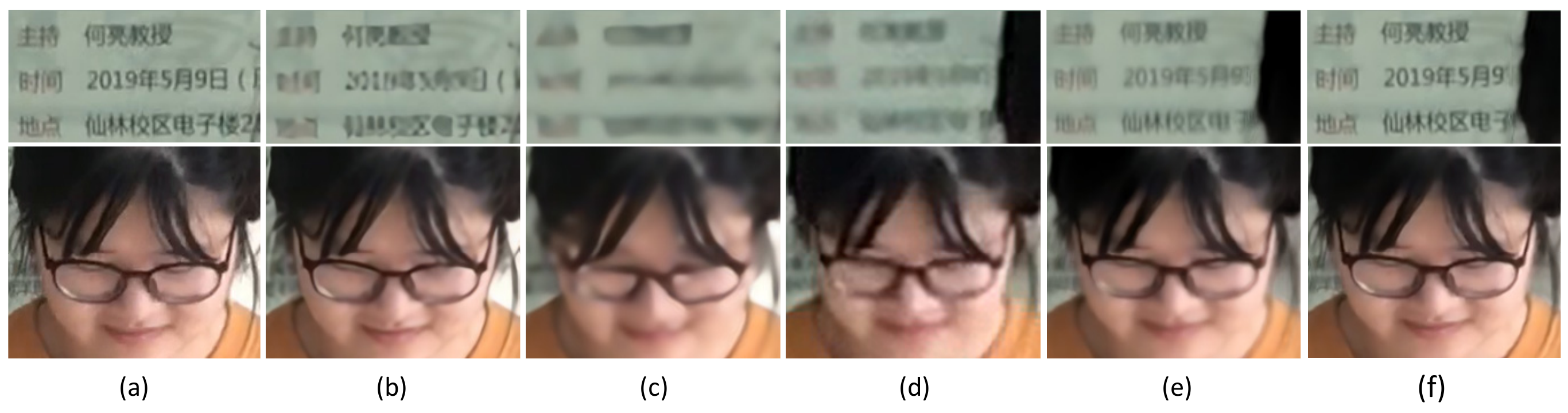}
    \caption{{\textbf{Resolution Gap Impact.}} (a) $I_{\tt ref}$, (b)  $I_{\tt ref}$ with 4$\times$ resolution downscaling and upscaling to original size by EDSR, (c)   $I_{\tt ref}$ with $8\times$ resolution downscaling and upscaling to original size by EDSR, (d) $I_{\tt LSR}$, (e) $Y$ with 4$\times$-Model, (f) $Y$ with 8$\times$-Model.}
    \label{fig:x4_vs_x8}
\end{figure*}

{\it Comparison Using Real Data:}  We choose a pair of Grasshopper3 cameras to perform more parallax studies due to its easy setup using industrial cameras. We use two Grasshopper3 GS3-U3-51S5C cameras with respective 20mm and 6mm lens installed. There is nearly 4$\times$ resolution gap between these two cameras, e.g., one is at {{2304$\times$2048, and the other one is at 576$\times$512}}. We fix the frame rate of the HSR-LFR camera at 240FPS and the frame rate of the LSR-HFR camera at 30FPS.  Viewing distance from the cameras to the scene is about 2 meters, and the baseline distance between these two cameras are adjusted at 5cm, 10cm, 15cm, 20cm and 25cm for a variety of parallax configurations. Figure~\ref{fig:camera_parallax} plots the reconstructed images at different baseline distances. As we can see, our system has reliable performance at a variety of parallax settings. Image quality can be enhanced noticeably with noise regularization, as shown in enlarged thumbnails in Figure~\ref{fig:camera_parallax}. And in the region with repeated patterns, the checkerboard, there are some ghosts on the results of CrossNet, but our network does not have this problem. Timer digits are over-smoothed by CrossNet, especially for the scenarios with larger baseline distance (e.g., 25cm), but ours still retain the sharp presentation.


{\bf Exposure time.}
The exposure time affects the number of arrival photons, thus having impacts on the image quality  for each snapshot and subsequent synthesis performance. Identical dual camera setup is used as in parallax study, but with the baseline distance fixed at 4.4cm. We fix the aperture sizes of the two cameras and set the ISO gain to automatic mode.

The exposure time of the HSR-LFR camera is fixed as default to record 30FPS reference video with 2304$\times$1920 resolution at 30FPS. For comparative studies, {{We set the exposure time of the LSR-HFR camera to 10ms}, 5ms, 2.5ms , 1ms  and 0.5ms respectively, to record video with  576$\times$480 resolution with 100FPS. Both captured and reconstructed images are shown in Fig.~\ref{fig:exposure_noise2}. From the captured LSR-HFR frames, we can see that the signal-to-noise ratio (SNR) of the images decrease greatly (Fig.~\ref{fig:exposure_noise2}(d)) as the decrease of the exposure time.}

%

%
%

{{Experiments reveal that CrossNet can remove some noise but the capacity is limited. This may be due to the smoothing effect of global convolutions applied, leading to blurred timing digits and jewelry contour (see Fig.~\ref{fig:exposure_noise2}(e)).
Our AWnet can effectively alleviate the noise (even with strong level) and maintain the sharpness, yielding much high-quality HSTR frames.}}
The noise induced by the lower SNR (with shorter exposure time), is greatly removed by our method (as illustrated in Fig.~\ref{fig:exposure_noise2}(d)-(e)), providing visually pleasant reconstruction with appealing spatial and temporal details. From these snapshots (and zoomed thumbnails), we can see that our system is robust and reliable to various exposure settings as well.

{\bf Motion blur.} Camera acquisition with insufficient temporal resolution  would introduce the motion blur as the zoomed regions of HSR-LFR frames in  Fig.~\ref{fig:motion_blur_new}(a). Our work is trying to leverage the LSR-HFR video with accurate motion details to resolve it. Towards this goal, dedicated FlowNet and FusionNet are devised to extract and aggregate information  (see Fig.~\ref{sfig:refSR}) for adaptive weighted synthesis. Optical flow extracted by the FlowNet is used to warp the $I_{\tt ref}$ followed by the dynamic filter and mask generated by the FusionNet to weigh the respective information from $I_{\tt ref} $ and $I_{\tt LSR}$ appropriately.

As shown in Fig.~\ref{fig:motion_blur_new}, motion blurs are clearly observed in  Fig.~\ref{fig:motion_blur_new}(a) around fast moving objects. Such effect is slightly reduced in Fig.~\ref{fig:motion_blur_new}(c) when flow information from LSR-HFR is utilized to warp the $I_{\tt ref}$.  Further improvement is achieved in Fig.~\ref{fig:motion_blur_new}(d) by utilizing the dynamic filter to restore the textures on blurred objects using information from the LSR-HFR input (see Fig.~\ref{fig:motion_blur_new}(b)(c)(d)). Sometimes, artifacts are induced due to over filtering, but subsequent adaptive mask in Fig.~\ref{fig:motion_blur_new}(e) will then intelligently combine pixels from $I_{\tt LSR\uparrow}$ and $I_{\tt ref}^{wk}$ for even better reconstruction shown in Fig.~\ref{fig:motion_blur_new}(f).

\textbf{Resolution Gap.}  An interesting observation is that our model trained with image pairs (see Section \ref{sec:train}) having $8\times$ resolution gap (noted as 8$\times$-Model) provides much better reconstruction quality subjectively (Fig.~\ref{fig:x4_vs_x8}(f)) compared to the model trained using image pairs with 4$\times$ resolution gap (noted as 4$\times$-Model) (Fig.~\ref{fig:x4_vs_x8}(e)). For both models, noise level is set with $\sigma^2$ = 0.01 for regularization.  For illustrative comparison, we downscale the $I_{\tt ref}$ to its $\frac{1}{16}$th  (e.g., $4\times$ downscaling at each spatial dimension) and $\frac{1}{64}$th (e.g., $8\times$ downscaling at each spatial dimension) sizes. Perceptually, a snapshot from the LSR-HFR camera in Fig.~\ref{fig:x4_vs_x8}(d) is close to $8\times$ downscaled $I_{\tt ref}$ in Fig.~\ref{fig:x4_vs_x8}(c), but  worse than  $4\times$ downscaled $I_{\tt ref}$ in Fig.~\ref{fig:x4_vs_x8}(b). Thus, when we use 4$\times$-Model, our network will evenly weigh information from $I_{\tt ref}$ and $I_{\tt LSR}$ yielding a smooth reconstruction with moderate quality in Fig.~\ref{fig:x4_vs_x8}(e); but for 8$\times$-Model, since the 8$\times$ downscaled version in training is with pretty bad quality, more weights will be given to $I_{\tt ref}$ in stationary areas and to $I_{\tt LSR}$ in motion areas for adaptive fusion synthesis, leading to much sharp details in reconstruction as shown in Fig.~\ref{fig:x4_vs_x8}(f). In other words, this is another example of adaptive weighting between the HSR-LFR and LSR-HFR camera inputs for final reconstruction quality improvement, where those weights can be regularized during training using sample pairs with different resolution gaps.

\section{Conclusion} \label{sec:conclusion}
A dual camera system is developed in this work for high spatiotemporal resolution video acquisition where one camera captures the HSR-LFR video, and the other one records the LSR-HFR video. An end-to-end learning framework, AWnet, is then proposed to learn the spatial and temporal information from both camera inputs, and drive the final appealing reconstruction by intelligently synthesizing the content from either HSR-LFR or LSR-HFR frame. Towards this goal, separable FlowNet and FusionNet are devised in our framework, to explicitly exploit the information from two cameras so as to derive the adaptive weighting functions for reconstruction synthesis.

Our system demonstrates the superior performance, in comparison to the existing works, such as the state-of-the-art CrossNet, PM, EDSR, and ToFlow-SR for super-resolution, and ToFlow-Intp for frame interpolation, showing noticeable gains both subjectively and objectively, using simulation data and camera captured real data.
ew
We also analyze various aspects of our system by breaking down its modular components, such as upscaling filter, reference structure, camera parallax, exposure time, etc. These studies pave the way for the application of our model to different scenarios.

In general, our system belongs to a hybrid camera or multi-camera category, even though our current emphasis is the production of video at both high spatial resolution and high frame rate. But this approach can be easily extended to view synthesis since the different viewpoints can be also generalized using flow representation. Another interesting avenue is to extend current RefSR mechanism in AWnet to include more cameras (e.g., $>2$) to enable the output video with more dimensional features, such as dynamic range (for low light imaging)~\cite{chen2018learning}, multi-spectra (beyond RGB), and depth (for 3D imaging)~\cite{fujii20193d}, etc.

Our AWnet could be further optimized towards the resource constrained embedded platform for broader applications using multi-camera equipped mobile phones, such as model compression~\cite{han2015deep}, simple yet effective network structure~\cite{howard2019searching}, architecture-driven software optimization~\cite{lane2016deepx}, etc.

\ifCLASSOPTIONcaptionsoff
  \newpage
\fi



%



\normalem
\bibliographystyle{ieee}
\bibliography{TPAMI-2019-09-0876-final}

\begin{thebibliography}{10}\itemsep=-1pt

\bibitem{StanfordLF}
The (new) stanford light field archive.
\newblock \url{http://http://lightfield.stanford.edu/lfs.html}.

\bibitem{baker2011database}
S.~Baker, D.~Scharstein, J.~Lewis, S.~Roth, M.~J. Black, and R.~Szeliski.
\newblock A database and evaluation methodology for optical flow.
\newblock {\em International Journal of Computer Vision}, 92(1):1--31, 2011.

\bibitem{DAIN}
W.~Bao, W.-S. Lai, C.~Ma, X.~Zhang, Z.~Gao, and M.-H. Yang.
\newblock Depth-aware video frame interpolation.
\newblock In {\em IEEE Conferene on Computer Vision and Pattern Recognition},
  2019.

\bibitem{bao2018memc}
W.~Bao, W.-S. Lai, X.~Zhang, Z.~Gao, and M.-H. Yang.
\newblock Memc-net: Motion estimation and motion compensation driven neural
  network for video interpolation and enhancement.
\newblock {\em arXiv preprint arXiv:1810.08768}, 2018.

\bibitem{bao2018high}
W.~Bao, X.~Zhang, L.~Chen, L.~Ding, and Z.~Gao.
\newblock High-order model and dynamic filtering for frame rate up-conversion.
\newblock {\em IEEE Transactions on Image Processing}, 27(8):3813--3826, 2018.

\bibitem{bay2008speeded}
H.~Bay, A.~Ess, T.~Tuytelaars, and L.~Van~Gool.
\newblock Speeded-up robust features (surf).
\newblock {\em Computer vision and image understanding}, 110(3):346--359, 2008.

\bibitem{boominathan2014improving}
V.~Boominathan, K.~Mitra, and A.~Veeraraghavan.
\newblock Improving resolution and depth-of-field of light field cameras using
  a hybrid imaging system.
\newblock In {\em 2014 IEEE International Conference on Computational
  Photography (ICCP)}, pages 1--10. IEEE, 2014.

\bibitem{brady2012multiscale}
D.~J. Brady, M.~E. Gehm, R.~A. Stack, D.~L. Marks, D.~S. Kittle, D.~R. Golish,
  E.~Vera, and S.~D. Feller.
\newblock Multiscale gigapixel photography.
\newblock {\em Nature}, 486(7403):386, 2012.

\bibitem{parallel_camera}
D.~J. Brady, W.~Pang, H.~Li, Z.~Ma, Y.~Tao, and X.~Cao.
\newblock Parallel cameras.
\newblock {\em Optica}, 5(2):127--137, Feb 2018.

\bibitem{burton1978thinking}
A.~Burton and J.~Radford.
\newblock {\em Thinking in Perspective: Critical Essays in the Study of Thought
  Processes}.
\newblock Psychology in progress. Methuen, 1978.

\bibitem{caballero2017real}
J.~Caballero, C.~Ledig, A.~Aitken, A.~Acosta, J.~Totz, Z.~Wang, and W.~Shi.
\newblock Real-time video super-resolution with spatio-temporal networks and
  motion compensation.
\newblock In {\em Proceedings of the IEEE Conference on Computer Vision and
  Pattern Recognition}, pages 4778--4787, 2017.

\bibitem{cao2016computational}
X.~Cao, T.~Yue, X.~Lin, S.~Lin, X.~Yuan, Q.~Dai, L.~Carin, and D.~J. Brady.
\newblock Computational snapshot multispectral cameras: Toward dynamic capture
  of the spectral world.
\newblock {\em IEEE Signal Processing Magazine}, 33(5):95--108, 2016.

\bibitem{chen2018learning}
C.~Chen, Q.~Chen, J.~Xu, and V.~Koltun.
\newblock Learning to see in the dark.
\newblock In {\em Proceedings of the IEEE Conference on Computer Vision and
  Pattern Recognition}, pages 3291--3300, 2018.

\bibitem{dong2015image}
C.~Dong, C.~C. Loy, K.~He, and X.~Tang.
\newblock Image super-resolution using deep convolutional networks.
\newblock {\em IEEE transactions on pattern analysis and machine intelligence},
  38(2):295--307, 2015.

\bibitem{eilertsen2017hdr}
G.~Eilertsen, J.~Kronander, G.~Denes, R.~K. Mantiuk, and J.~Unger.
\newblock Hdr image reconstruction from a single exposure using deep cnns.
\newblock {\em ACM Transactions on Graphics (TOG)}, 36(6):1--15, 2017.

\bibitem{eilertsen2019single}
G.~Eilertsen, R.~K. Mantiuk, and J.~Unger.
\newblock Single-frame regularization for temporally stable cnns.
\newblock In {\em Proceedings of the IEEE Conference on Computer Vision and
  Pattern Recognition}, pages 11176--11185, 2019.

\bibitem{freeman2002example}
W.~T. Freeman, T.~R. Jones, and E.~C. Pasztor.
\newblock Example-based super-resolution.
\newblock {\em IEEE Computer graphics and Applications}, (2):56--65, 2002.

\bibitem{fujii20193d}
T.~Fujii.
\newblock 3d image processing--from capture to display--.
\newblock {\em Electronic Imaging}, 2019(3):625--1, 2019.

\bibitem{gu2019blind}
J.~Gu, H.~Lu, W.~Zuo, and C.~Dong.
\newblock Blind super-resolution with iterative kernel correction.
\newblock In {\em Proceedings of the IEEE conference on computer vision and
  pattern recognition}, pages 1604--1613, 2019.

\bibitem{han2015deep}
S.~Han, H.~Mao, and W.~J. Dally.
\newblock Deep compression: Compressing deep neural networks with pruning,
  trained quantization and huffman coding.
\newblock {\em arXiv preprint arXiv:1510.00149}, 2015.

\bibitem{howard2019searching}
A.~Howard, M.~Sandler, G.~Chu, L.-C. Chen, B.~Chen, M.~Tan, W.~Wang, Y.~Zhu,
  R.~Pang, V.~Vasudevan, et~al.
\newblock Searching for mobilenetv3.
\newblock {\em arXiv preprint arXiv:1905.02244}, 2019.

\bibitem{FlowNet2}
E.~Ilg, N.~Mayer, T.~Saikia, M.~Keuper, A.~Dosovitskiy, and T.~Brox.
\newblock Flownet 2.0: Evolution of optical flow estimation with deep networks.
\newblock In {\em IEEE Conference on Computer Vision and Pattern Recognition
  (CVPR)}, Jul 2017.

\bibitem{jaderberg2015spatial}
M.~Jaderberg, K.~Simonyan, A.~Zisserman, et~al.
\newblock Spatial transformer networks.
\newblock In {\em Advances in neural information processing systems}, pages
  2017--2025, 2015.

\bibitem{jia2016dynamic}
X.~Jia, B.~De~Brabandere, T.~Tuytelaars, and L.~V. Gool.
\newblock Dynamic filter networks.
\newblock In {\em Advances in Neural Information Processing Systems}, pages
  667--675, 2016.

\bibitem{jiang2018super}
H.~Jiang, D.~Sun, V.~Jampani, M.-H. Yang, E.~Learned-Miller, and J.~Kautz.
\newblock Super slomo: High quality estimation of multiple intermediate frames
  for video interpolation.
\newblock In {\em Proceedings of the IEEE Conference on Computer Vision and
  Pattern Recognition}, pages 9000--9008, 2018.

\bibitem{jin2019learning}
M.~Jin, Z.~Hu, and P.~Favaro.
\newblock Learning to extract flawless slow motion from blurry videos.
\newblock In {\em Proceedings of the IEEE Conference on Computer Vision and
  Pattern Recognition}, pages 8112--8121, 2019.

\bibitem{jo2018deep}
Y.~Jo, S.~Wug~Oh, J.~Kang, and S.~Joo~Kim.
\newblock Deep video super-resolution network using dynamic upsampling filters
  without explicit motion compensation.
\newblock In {\em Proceedings of the IEEE Conference on Computer Vision and
  Pattern Recognition}, pages 3224--3232, 2018.

\bibitem{kim2016accurate}
J.~Kim, J.~Kwon~Lee, and K.~Mu~Lee.
\newblock Accurate image super-resolution using very deep convolutional
  networks.
\newblock In {\em Proceedings of the IEEE conference on computer vision and
  pattern recognition}, pages 1646--1654, 2016.

\bibitem{kingma2014adam}
D.~P. Kingma and J.~Ba.
\newblock Adam: A method for stochastic optimization.
\newblock {\em arXiv preprint arXiv:1412.6980}, 2014.

\bibitem{lane2016deepx}
N.~D. Lane, S.~Bhattacharya, P.~Georgiev, C.~Forlivesi, L.~Jiao, L.~Qendro, and
  F.~Kawsar.
\newblock Deepx: A software accelerator for low-power deep learning inference
  on mobile devices.
\newblock In {\em Proceedings of the 15th International Conference on
  Information Processing in Sensor Networks}, page~23. IEEE Press, 2016.

\bibitem{ledig2017photo}
C.~Ledig, L.~Theis, F.~Husz{\'a}r, J.~Caballero, A.~Cunningham, A.~Acosta,
  A.~Aitken, A.~Tejani, J.~Totz, Z.~Wang, et~al.
\newblock Photo-realistic single image super-resolution using a generative
  adversarial network.
\newblock In {\em Proceedings of the IEEE conference on computer vision and
  pattern recognition}, pages 4681--4690, 2017.

\bibitem{lowlight_imaging}
F.~Li.
\newblock {\em A HYBRID CAMERA SYSTEM FOR LOW-LIGHT IMAGING}.
\newblock PhD thesis, University of Delaware, 2011.

\bibitem{lim2017enhanced}
B.~Lim, S.~Son, H.~Kim, S.~Nah, and K.~Mu~Lee.
\newblock Enhanced deep residual networks for single image super-resolution.
\newblock In {\em Proceedings of the IEEE Conference on Computer Vision and
  Pattern Recognition Workshops}, pages 136--144, 2017.

\bibitem{liu2013bundled}
S.~Liu, L.~Yuan, P.~Tan, and J.~Sun.
\newblock Bundled camera paths for video stabilization.
\newblock {\em ACM Transactions on Graphics (TOG)}, 32(4):78, 2013.

\bibitem{lu2019high}
S.~Lu.
\newblock High-speed video from asynchronous camera array.
\newblock In {\em 2019 IEEE Winter Conference on Applications of Computer
  Vision (WACV)}, pages 2196--2205. IEEE, 2019.

\bibitem{menze2015object}
M.~Menze and A.~Geiger.
\newblock Object scene flow for autonomous vehicles.
\newblock In {\em Proceedings of the IEEE Conference on Computer Vision and
  Pattern Recognition}, pages 3061--3070, 2015.

\bibitem{niklaus2018context}
S.~Niklaus and F.~Liu.
\newblock Context-aware synthesis for video frame interpolation.
\newblock In {\em Proceedings of the IEEE Conference on Computer Vision and
  Pattern Recognition}, pages 1701--1710, 2018.

\bibitem{noh2017regularizing}
H.~Noh, T.~You, J.~Mun, and B.~Han.
\newblock Regularizing deep neural networks by noise: Its interpretation and
  optimization.
\newblock In {\em Advances in Neural Information Processing Systems}, pages
  5109--5118, 2017.

\bibitem{ronneberger2015u}
O.~Ronneberger, P.~Fischer, and T.~Brox.
\newblock U-net: Convolutional networks for biomedical image segmentation.
\newblock In {\em International Conference on Medical image computing and
  computer-assisted intervention}, pages 234--241. Springer, 2015.

\bibitem{sajjadi2018frame}
M.~S. Sajjadi, R.~Vemulapalli, and M.~Brown.
\newblock Frame-recurrent video super-resolution.
\newblock In {\em Proceedings of the IEEE Conference on Computer Vision and
  Pattern Recognition}, pages 6626--6634, 2018.

\bibitem{srinivasan2017learning}
P.~P. Srinivasan, T.~Wang, A.~Sreelal, R.~Ramamoorthi, and R.~Ng.
\newblock Learning to synthesize a 4d rgbd light field from a single image.
\newblock In {\em Proceedings of the IEEE International Conference on Computer
  Vision}, pages 2243--2251, 2017.

\bibitem{sun2018pwc}
D.~Sun, X.~Yang, M.-Y. Liu, and J.~Kautz.
\newblock Pwc-net: Cnns for optical flow using pyramid, warping, and cost
  volume.
\newblock In {\em Proceedings of the IEEE Conference on Computer Vision and
  Pattern Recognition}, pages 8934--8943, 2018.

\bibitem{wang2017light}
T.-C. Wang, J.-Y. Zhu, N.~K. Kalantari, A.~A. Efros, and R.~Ramamoorthi.
\newblock Light field video capture using a learning-based hybrid imaging
  system.
\newblock {\em ACM Transactions on Graphics (TOG)}, 36(4):133, 2017.

\bibitem{wang2004image}
Z.~Wang, A.~C. Bovik, H.~R. Sheikh, E.~P. Simoncelli, et~al.
\newblock Image quality assessment: from error visibility to structural
  similarity.
\newblock {\em IEEE transactions on image processing}, 13(4):600--612, 2004.

\bibitem{wu2017light}
G.~Wu, B.~Masia, A.~Jarabo, Y.~Zhang, L.~Wang, Q.~Dai, T.~Chai, and Y.~Liu.
\newblock Light field image processing: An overview.
\newblock {\em IEEE Journal of Selected Topics in Signal Processing},
  11(7):926--954, 2017.

\bibitem{xie2019self}
Q.~Xie, E.~Hovy, M.-T. Luong, and Q.~V. Le.
\newblock Self-training with noisy student improves imagenet classification.
\newblock {\em arXiv preprint arXiv:1911.04252}, 2019.

\bibitem{xue2019video}
T.~Xue, B.~Chen, J.~Wu, D.~Wei, and W.~T. Freeman.
\newblock Video enhancement with task-oriented flow.
\newblock {\em International Journal of Computer Vision (IJCV)},
  127(8):1106--1125, 2019.

\bibitem{zhang2019deep}
K.~Zhang, W.~Zuo, and L.~Zhang.
\newblock Deep plug-and-play super-resolution for arbitrary blur kernels.
\newblock In {\em Proceedings of the IEEE Conference on Computer Vision and
  Pattern Recognition}, pages 1671--1681, 2019.

\bibitem{zheng2017combining}
H.~Zheng, M.~Guo, H.~Wang, Y.~Liu, and L.~Fang.
\newblock Combining exemplar-based approach and learning-based approach for
  light field super-resolution using a hybrid imaging system.
\newblock In {\em Proceedings of the IEEE International Conference on Computer
  Vision}, pages 2481--2486, 2017.

\bibitem{zheng2017learning}
H.~Zheng, M.~Ji, H.~Wang, Y.~Liu, and L.~Fang.
\newblock Learning cross-scale correspondence and patch-based synthesis for
  reference-based super-resolution.
\newblock In {\em BMVC}, 2017.

\bibitem{zheng2018crossnet}
H.~Zheng, M.~Ji, H.~Wang, Y.~Liu, and L.~Fang.
\newblock Crossnet: An end-to-end reference-based super resolution network
  using cross-scale warping.
\newblock In {\em Proceedings of the European Conference on Computer Vision
  (ECCV)}, pages 88--104, 2018.

\bibitem{zheng2016improving}
S.~Zheng, Y.~Song, T.~Leung, and I.~Goodfellow.
\newblock Improving the robustness of deep neural networks via stability
  training.
\newblock In {\em Proceedings of the ieee conference on computer vision and
  pattern recognition}, pages 4480--4488, 2016.

\end{thebibliography}

%



\clearpage
\begin{IEEEbiography}
[{\includegraphics[width=1in,height=1.25in,clip,keepaspectratio]{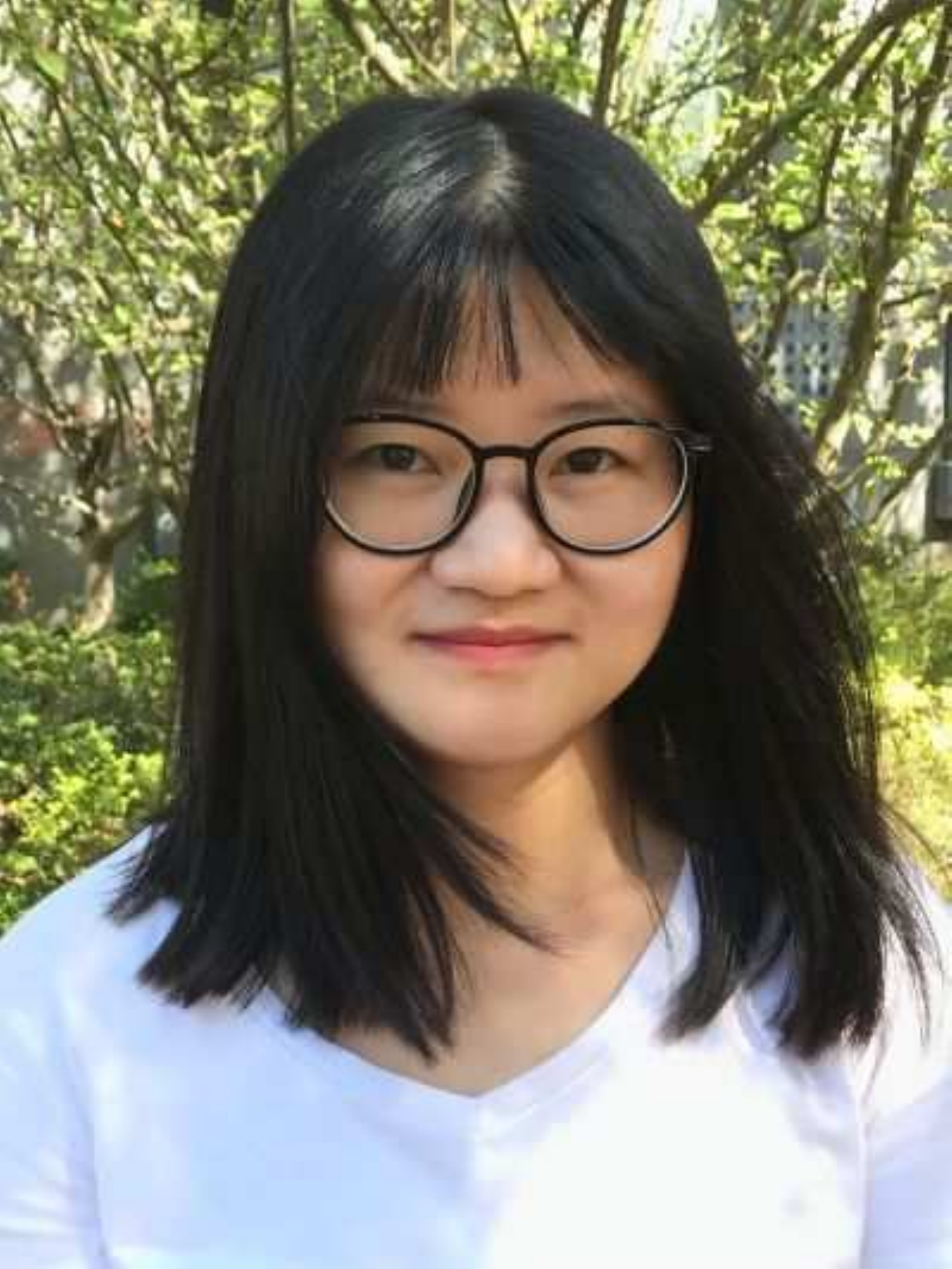}}]
{Ming Cheng}
is a Ph.D. candidate of Electrical Engineering with the Institute of Image Communication and Network Engineering, Shanghai Jiao Tong University, Shanghai, China.
She received the B.S. degree in the University of Electronic Science and Technology of China, China, in 2016.
Her research interests include computer vision, video processing and multi-cameras.
\end{IEEEbiography}
\begin{IEEEbiography}
[{\includegraphics[width=1in,height=1.25in,clip,keepaspectratio]{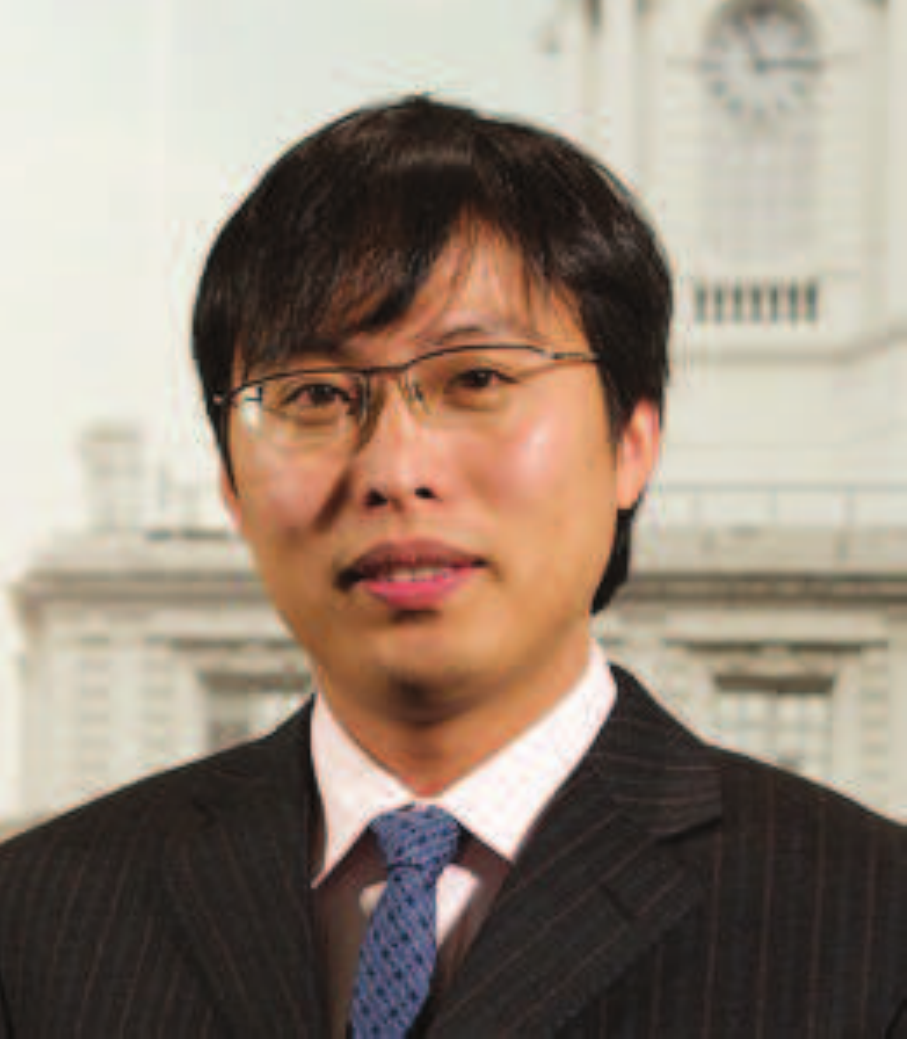}}]
{Zhan Ma}
(SM'19) is now on the faculty of Electronic Science and Engineering School, Nanjing University, Jiangsu, 210093, China.
He received the B.S. and M.S. from Huazhong University of Science and Technology (HUST), Wuhan, China, in 2004 and 2006 respectively, and the Ph.D. degree from the New York University, New York, in 2011.
From 2011 to 2014, he has been with Samsung Research America, Dallas TX, and  Futurewei Technologies, Inc., Santa Clara, CA, respectively. His current research focuses on the next-generation video coding, energy-efficient communication, gigapixel streaming and deep learning. He is a co-recipient of 2018 ACM SIGCOMM Student Research Competition Finalist, 2018 PCM Best Paper Finalist, and 2019 IEEE Broadcast Technology Society Best Paper Award.
\end{IEEEbiography}
\begin{IEEEbiography}
[{\includegraphics[width=1in,height=1.25in,clip,keepaspectratio]{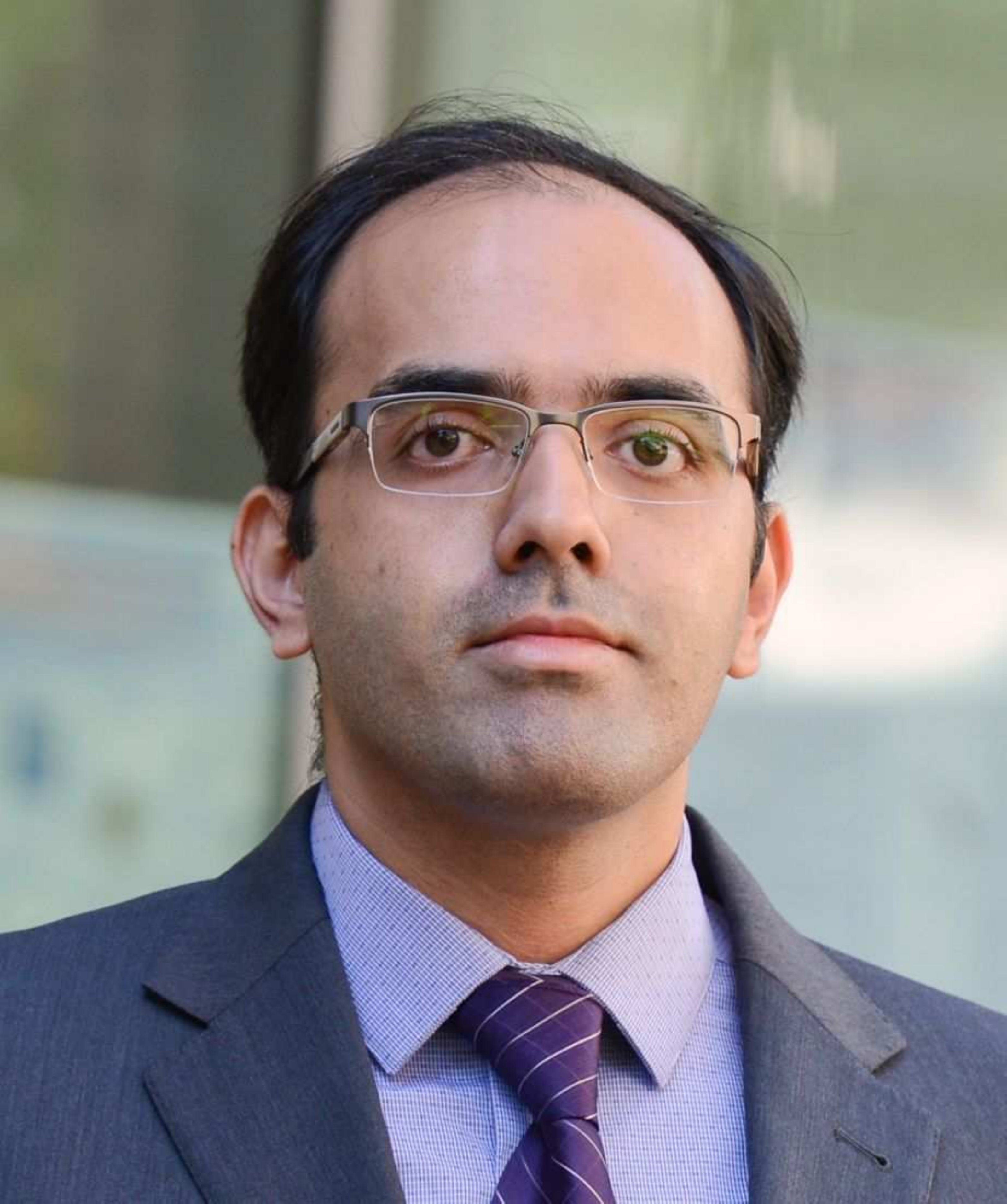}}]
{M. Salman Asif}
is currently an Assistant Professor in the Department of Electrical and Computer Engineering at the University of California, Riverside.
Prior to joining UC Riverside, he was a postdoctoral research associate in the DSP group at Rice University. Before that he briefly worked as a Research Engineer at Samsung Research America, Dallas. He received the Ph.D. at the Georgia Institute of Technology under the supervision of Justin Romberg.
His research interests broadly lie in the areas of information processing and computational sensing with applications in signal processing, machine learning, and computational imaging.
\end{IEEEbiography}
\begin{IEEEbiography}
[{\includegraphics[width=1in,height=1.25in,clip,keepaspectratio]{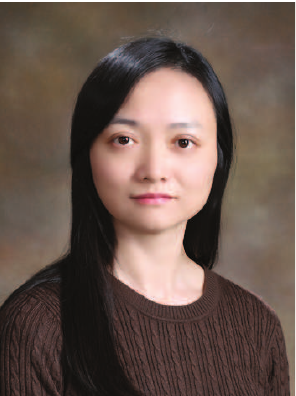}}]
{Yiling Xu}
is a full researcher of School of Electronic Information and Electronic Engineering, Shanghai Jiao Tong University, Shanghai, 200145, China.
She received the B.S., M.S. and Ph.D. from the University of Electronic Science and Technology of China, China, in 1999, 2001 and 2004 respectively. From 2004 to 2013, she was with Multimedia Communication Research Institute of Samsung Electronics Inc, Korea.
Her main research interests include architecture design for next generation multimedia systems, dynamic data encapsulation, adaptive cross layer design, dynamic adaption for heterogenous networks and N-screen content presentation.
\end{IEEEbiography}
\begin{IEEEbiography}
[{\includegraphics[width=1in,height=1.25in,clip,keepaspectratio]{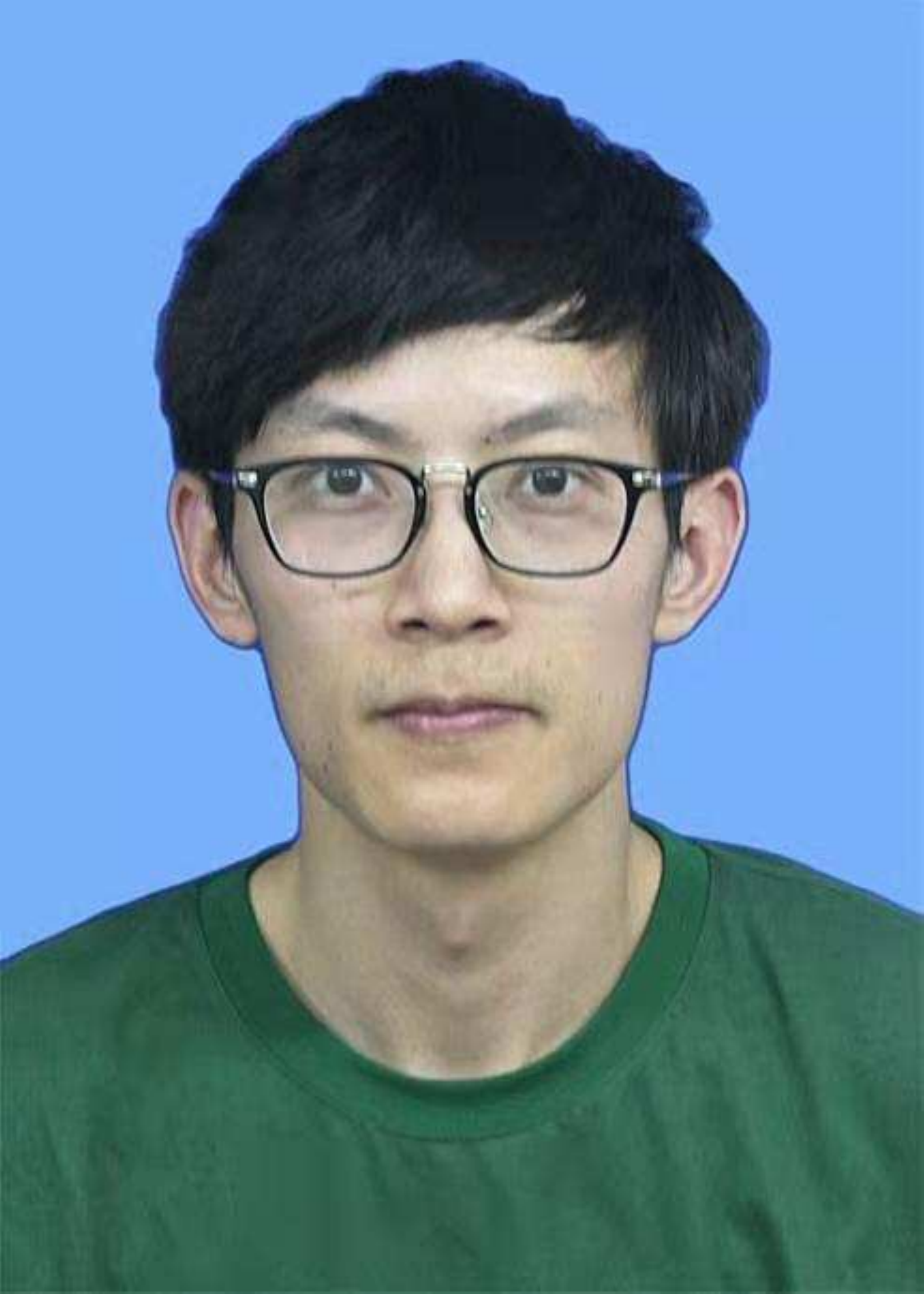}}]
{Haojie Liu}
is a Ph.D candidate in School of Electronic Science and Engineering, Nanjing University, Nanjing, China. He received the B.S degree in Nanjing University in 2016. His research interests includes video communication and processing, machine learning and computer vision.
\end{IEEEbiography}
\begin{IEEEbiography}
[{\includegraphics[width=1in,height=1.25in,clip,keepaspectratio]{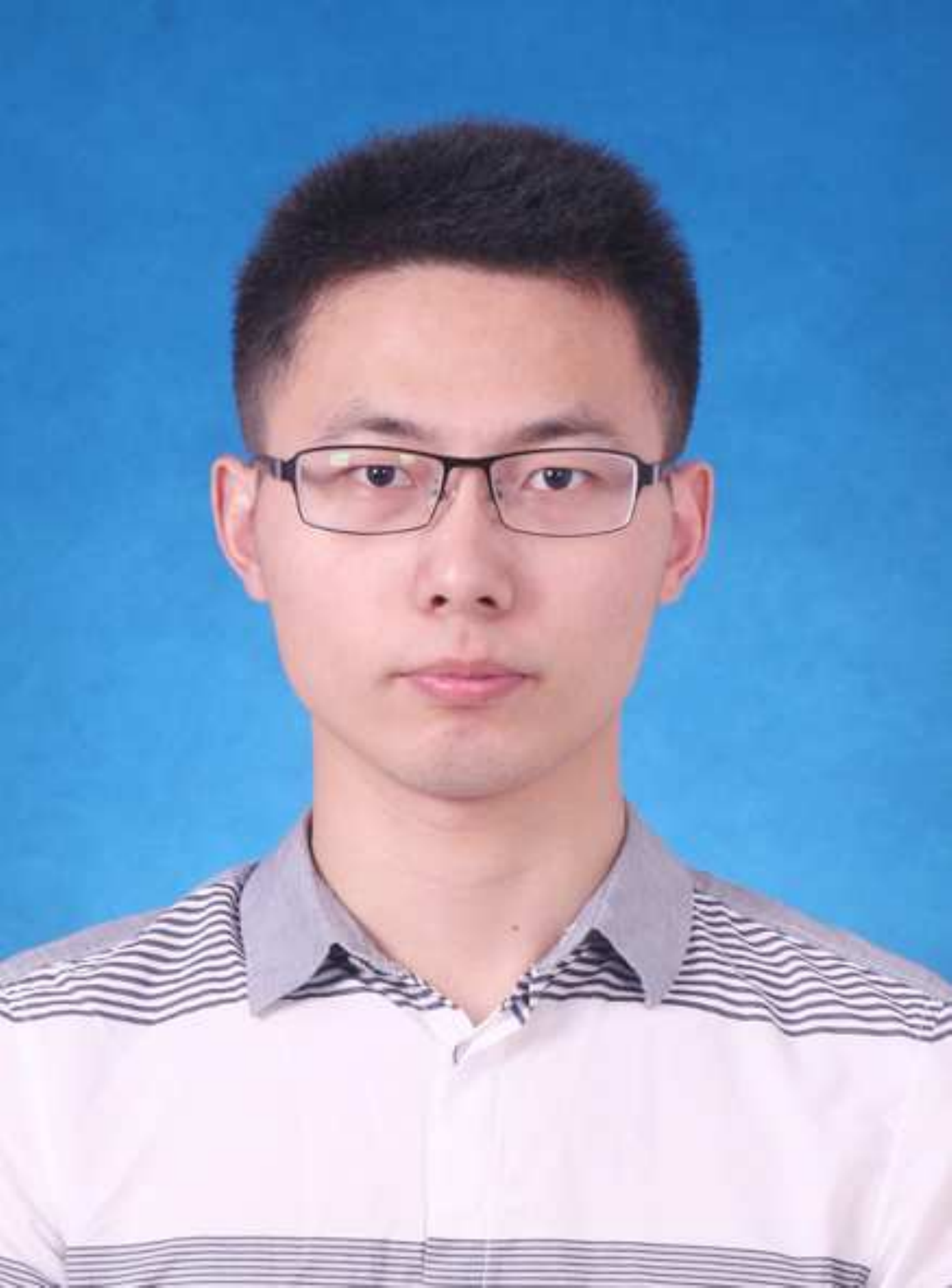}}]
{Wenbo Bao}
is a Ph.D. candidate of Electrical Engineering with the Institute of Image Communication and Network Engineering, Shanghai Jiao Tong University, Shanghai, China.
He received the B.S. degree in Electronic Information Engineering from Huazhong University of Science and Technology, Hubei, China, in 2014.
His research interests include computer vision, machine learning, and video processing.
\end{IEEEbiography}
\begin{IEEEbiography}
[{\includegraphics[width=1in,height=1.25in,clip,keepaspectratio]{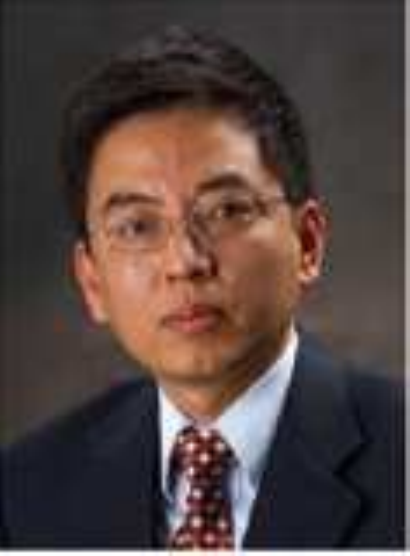}}]
{Jun Sun}
is currently a professor and Ph.D. advisor of Shanghai Jiao Tong University.
He received his B.S. in 1989 from University of Electronic Sciences and technology of China, Chengdu, China, and a Ph.D. degree in 1995 from Shanghai Jiao Tong University, all in electrical engineering.
In 1996, he was elected as the member of HDTV Technical Executive Experts Group (TEEG) of China. Since then, he has been acting as one of the main technical experts for the Chinese government in the field of digital television and multimedia communications. In the past five years, he has been responsible for several national projects in DTV and IPTV fields. He has published over 50 technical papers in the area of digital television and multimedia communications and received 2nd Prize of National Sci. \& Tech. Development Award in 2003, 2008.
His research interests include digital television, image communication, and video encoding.
\end{IEEEbiography}

\end{document}